\documentclass[pss]{wiley2sp} 
\usepackage{amsmath} \usepackage{bm} \usepackage[percent]{overpic}
\usepackage{threeparttable,booktabs}
\usepackage{xcolor}
\begin{document}

\title{Morphology, ordering, stability, and electronic structure of carbon-doped
hexagonal boron nitride}

\author{ Agnieszka Jamr\'{o}z, Jacek A. Majewski}

\mail{e-mail \textsf{agnieszka.jamroz@fuw.edu.pl}}

\institute{Faculty of Physics, University of Warsaw,
Pasteura 5, 02-093 Warsaw, Poland\\
}


\abstract{\bf
 We present theoretical studies of morphology, stability, and
electronic structure of monolayer hexagonal CBN alloys with rich content of h-BN
and carbon concentration not exceeding 50\%. Our studies are based on the bond
order type of the valence force field to account for the interactions between
atomic constituents and Monte Carlo method with Metropolis algorithm to
establish equilibrium distribution of atoms over the lattice. We find out that
the phase separation into graphene and h-BN domains occurs in the majority of
growth conditions. Only in N-rich growth conditions, it is possible to obtain
quasi uniform distribution of carbon atoms over boron sublattice. We predict
also that the energy gap in stoichiometric C$_x$(BN)$_{1-x}$ alloys exhibits
extremely strong bowing.}

\maketitle  

\section{Introduction}

\begin{figure}[h!tb] \includegraphics[width=0.45 \textwidth]{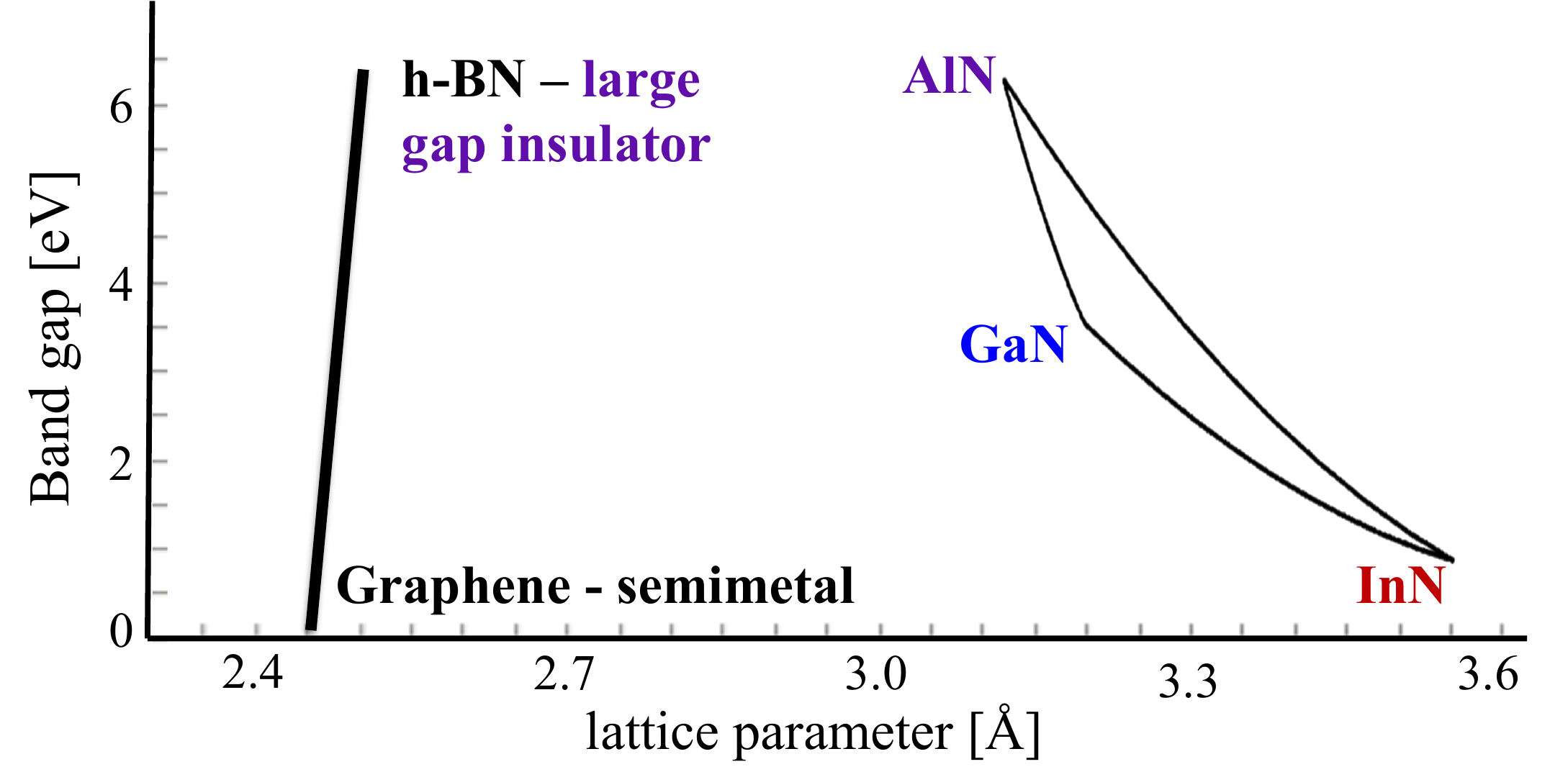}
\caption{Comparison of lattice constants and band gaps for nitride materials
(GaN, InN, AlN) and h-BN-graphene.
	}
	\label{Fig1} \end{figure}

Two-dimensional, layered, atomically thin systems play important role in
contemporary materials science. Hexagonal layered boron-carbide-nitrides, i.e.,
nanosystems with boron, nitrogen, and carbon atoms distributed over the
honeycomb lattice, constitute rich and interesting family of the 2D materials
\cite{BCN-BOOK,CBN-materials,BxCyNz}. The two limiting cases of such systems are
graphene and monolayer of hexagonal boron nitride, and these systems can be
considered as building blocks of layered CBN alloys. Depending on the relative
concentrations of C, B, and N, they might be on one hand referred to as boron-
and/or nitrogen-doped graphene (sometimes called B-graphene, N-graphene, or,
BN-graphene, respectively), or on the other hand, as carbon doped hexagonal
boron nitride (h-BN), very often indicated as C-hBN, or C$_x$(BN)$_{1-x}$. In
addition, these layered CBN alloys can be combined into vertically stacked
heterostructures \cite{PRB-Sachs}, in a similar manner to fairly intensively
studied nowadays the so-called van der Waals (vdW) heterostructures of graphene
and h-BN. Such layered CBN alloys could definitely enrich the families of both
vdW and also lateral heterostructures and are definitely worth of basic
investigations. As it was mentioned above, the 2D layered CBN system can be
considered as an alloy of graphene and h-BN. These two ingredients of the alloy
are nearly lattice matched and exhibit the lattice mismatch of only 1\% (the
lateral lattice constants of graphene and h-BN are equal to 2.46 \AA
\cite{agrapph} and 2.504 \AA \cite{ahBN}, respectively). Simultaneously, their
fundamental band gaps differ by $\sim$6\,eV, ranging from semimetal to high band
gap insulator. Therefore, the layered CBN alloys should allow for band gap
tuning within this range.  This uniqueness of CBN layered alloys is emphasized
in Figure~\ref{Fig1}, where an illustrative comparison with the family of
III-Nitrides (GaN, AlN, and InN as border materials) is given. It is, therefore,
very natural that the layered CBN alloys are considered as obvious and promising
candidates for numerous applications not only in the area of functional
electronic and optoelectronic devices \cite{EL-OPT}, but also in many other
fields such as carbon capture \cite{CO2capt}, high energy density
supercapacitors \cite{HEDS}, Li-Ion batteries \cite{Li-bat}, metal-free
photoredox catalysis \cite{REDOX}, effective and highly selective H$_{2}$
separation membranes \cite{MEMBRANE}, and many others.

In this paper, we focus on h-BN rich CBN alloys (i.e., with carbon content not
exceeding 50\%) and perform extensive theoretical studies of their
morphology, ordering, stability, and electronic structure. These studies should shed 
light on the physical mechanisms leading to
the synthesis of alloys with various phases and ordering patterns observed
experimentally. Before we present the details of the employed methodology and
the results of the studies, we would like to survey very briefly the basis
experimental facts about the studied layered CBN alloys. \newline

There are numerous approaches to synthesize carbon-containing hexagonal boron
nitride, including CVD, MOVPE, or post-synthesis implantation of carbon into
pristine h-BN layer. The possibility to synthesize one monolayer of hexagonal
boron nitride was proved few years ago \cite{hBN-0,hBN-1}. In this pioneering
studies, monolayers of h-BN were formed on the Ni(111), Pd(111), and Pt(111)
surfaces, and the electronic dispersion relations of the monolayer were measured
\cite{hBN-0,hBN-1}. Later on, the h-BN monolayer was fabricated by Chemical
Vapor Deposition (CVD) and Metal Organic CVD (MOCVD) techniques
\cite{hBN-synth1,hBN-CVD,hBN-MOCVD}. Employing various growth techniques,
synthesis of h-CBN graphetic but ternary monolayer was also successfully
performed. For example, to synthesize h-CBN, the methane and ammonia borane were
introduced at the same time to act as precursors for carbon and BN, respectively
\cite{CBN-materials}, or in other growth process the bis-BN cyclohexane,
B$_2$N$_2$C$_2$H$_{12}$, was utilized for the synthesis of h-CBN \cite{h-BCN1}.
Typically, the atomic ratio of B, C, and N can be tuned to some extent by
changing the growth conditions, which leads to h-CBN samples of various homogeneity and
morphology \cite{h-BCN1}. Electron-beam mediated post-synthesis doping 
of boron-nitride nanostructures with carbon atoms was also demonstrated 
\cite{Nature464,ACSNano11}.  
Generally, from experimental studies
the following picture emerges: (i) the separation of phases occurs typically in
growth conditions close to equilibrium; (ii) however, the ability in h-CBN
alloys to form homophilic (C-C) as well as heterophilic (C-B, C-N, B-N) bonds
generates a rich variety of polymorphic structures, making the precise control
of their chemical stoichiometry and geometry formidable \cite{Natcom}.

As a consequence, according to the authors of Ref. \cite{Natcom}, detailed
experimental insights into phase separation and ordering in such alloys are
currently lacking. This strongly suggests the need of employing reliable
theoretical modelling to increase understanding of these nano-structures, 
and some theoretical studies of the stability and electronic structure of the h-CBN 
alloys were already performed, both for ordered prototypes of alloys 
\cite{rev2_4,PRB73,PRB84,rev1_1,rev1_2,rev2_2,rev2_3,JpnJApplPhys} 
and taking configurational disorder into account
\cite{PRB79,PRB95,JAComp708}. 

In Ref. \cite{PRB79} phase stability for monolayer BNC ternary system was examined 
by Monte Carlo simulations and the cluster expansion effective Hamiltonian with parameters 
determined from the density functional theory (DFT) calculations. The studies concluded 
that all possible atomic arrangements exhibited positive formation energies, 
indicating phase separation into
monolayer BN and graphene. The authors also concluded that only C-C and B-N bonds
would be energetically favorable in the alloy, whereas all others, i.e., C-N, B-C,
B-B, and N-N would be not. Also the optical absorbance and possibility of band gap
engineering, phase separation, and composition fluctuations of C$_x$(BN)$_{1-x}$
alloys were investigated \cite{PRB95}, where the DFT calculations combined with the 
generalized quasi-chemical approximation (GQCA) to account for disorder effects were employed. 
In that study the tendency to separation into h-BN and graphene was found to be strong, 
however, some mixing was allowed. 
In Ref. \cite{JAComp708}, an approach combining the DFT calculations 
and Monte Carlo was used to investigate the influence of interface geometry on phase 
stability and band gap engineering in boron nitride substituted graphene. 
The Monte Carlo sampling was performed with the model Hamiltonian defined 
on a bond basis with bond energies extracted out from DFT calculations. 
The authors came to similar conclusions as in the papers \cite{PRB79,PRB95}, 
namely that the stability and properties of CBN alloys would be governed by the fact 
that C-C and B-N bonds are energetically more favorable than other bonds in the systems. 

It is worth to note, that in all the studies of CBN alloys mentioned above, 
the DFT calculations that were used either directly to find out the physical 
mechanisms determining the stability of CBN systems or were used 
as a kind of intermediate procedure to determine 
parameters of model Hamiltonians utilized for further investigation of systems' 
equilibrium configuration with 
Monte Carlo method dealt with rather small alloy structures, containing 
typically up to few dozens or sometimes hundreds of atoms \cite{rev2_3}. 
In our methodology, we follow the
direct approach, albeit empirical in nature, which allows us to treat systems
consisting of thousands of atoms, either with or without periodic boundary
conditions. This approach (named VFF-MC) is based on the valence force field (VFF) 
to calculate the total energy of alloys and Monte Carlo method to determine the equilibrium
distribution of atoms over the lattice. We have used successfully the VFF-MC scheme 
(with Keating type of VFF) to investigate ordering issues in III-nitride ternary and 
quaternary alloys \cite{Michal}. We have also investigated chemical order on fixed 
hexagonal lattice in B-, N-, and BN-doped graphene monolayer \cite{AJ} with VFF of 
bond order Tersoff's potential \cite{e[2.1]}. This potential has successfully history of 
implementations in huge number of studies dealing with systems composed of carbon, 
nitrogen, and also boron. In our previous studies \cite{AJ}, we considered carbon reach 
CBN alloys with concentration of dopants not exceeding 50\%. In the present study, 
we address the h-BN rich side of mono-layer CBN alloys, with carbon part below 50\%. This
contributes to the hotly discussed issue of carbon doped nitrides. In addition,
we have refined our VFF-MC approach: (i) allowing for lattice relaxation, and (ii)
extending scheme for electronic structure calculations of alloys.

The paper is organized as follows. In Section~\ref{Methods-Tersoff} we describe
first the specific implementation of VFF Tersoff's potential and then MC-VFF
simulation scheme in Section~\ref{Methods-VFF-MC}). 
Some kind of validation of chosen Tersoff's potential is given 
in Section ~\ref{Methods-tests}. The main results of this paper are
discussed in Section~\ref{RESULTS}, and finally the paper is summarized in
Section~\ref{Summary}.

\section{Methods} 
\subsection{Tersoff potential for simulating carbon doped
boron nitride systems} 
\label{Methods-Tersoff} 

The CBN alloys as systems lacking long-range order and exhibiting short range 
order in turn are extremely computationally demanding for any atomistic scheme. 
To obtain reliable theoretical predictions of morphology and structural properties 
of these alloys, one needs typically to perform calculations for thousands of atoms, 
which is out of scope of presently so widely used {\it ab initio} methods. 
Therefore, the attractive alternative is to employ the VFF type of potential that makes 
the calculations feasible. Our previous studies for B
and N doped graphene have confirmed that the semi-empirical Tersoff potential
(T-P) \cite{e[2.1]}, which belongs to the family of bond-order potentials and is
fairly widely exploited in computational materials science, provides reasonable 
description of such systems. In contrast to classical two-body interatomic potentials (e.g.,
Lennard-Jones, Keating), the interaction strength between two atoms in the T-P
depends not only on their mutual distance, but also on the closest environment
(coordination number and types of neighboring atoms). This feature of the T-P
allows even for the correct description of chemical reactions.

In the present work, we use T-P for multicomponent systems following the formulation
of Kroll \cite{Kroll}. The total energy of the system is described as a sum over
two center interatomic interactions $V_{ij}$ between atomic pairs $i$ and $j$
separated by the distance $r_{ij}$

\begin{equation} \label{equation:Tersoff1} V_{ij} = f_C (r_{ij}) \left(
f_R(r_{ij}) + b_{ij} f_A(r_{ij})\right), 
\end{equation}

\noindent where $f_R$ and $f_A$ constitute repulsive and attractive interaction,
respectively, and $f_C$ is cut-off function, just to ensure potential's short
range. $R_{ij}$ and $S_{ij}$ are typically chosen to include only the first
shell of neighbors \footnote{In present research we used: $R_{CC} = 1.8$ \AA, $R_{BB} =
R_{NN} = R_{BN} = 1.9$ \AA, $R_{CB} = R_{CN} = 1.85$ \AA, $S_{CC} = S_{BB} =
S_{NN} = 2.1$ \AA,  $R_{BN} =2.0$ \AA, $S_{CB} = S_{CBN} = 2.05$ \AA.}

\begin{equation} f_A (r_{ij}) = A_{ij} e^{-\lambda_{ij} r_{ij}},\>\>\>\>\> f_R
(r_{ij}) = B_{ij} e^{-\mu_{ij} r_{ij}}, \end{equation}

\begin{equation} f_C(r_{ij})= \left\{ \begin{array}{l}1 ,\>\>\> r_{ij} <
R_{ij}\\ \frac{1}{2} + \frac{1}{2} cos\left( \frac {\pi (r_{ij}-
R_{ij})}{(S_{ij} - R_{ij})} \right),\>\>\> R_{ij}< r_{ij} < S_{ij}. \\0,\>\>\>
r_{ij} > S_{ij} \end{array} \right. \end{equation}

The bond-order term $b_{ij}$ describes the influence of neighborhood on the
binding strength of a pair of atoms by modifying the attractive part of the
potential

\begin{equation} \label{eq:beta_ijk} b_{ij} = \chi_{ij} \left( 1 +
\zeta_{ij}^{n_{i}} \right)^{-\frac{1}{2n_i}}, \end{equation}

\begin{equation} \zeta_{ij} = \sum_{k \neq i,j} \>f_C(r_{ik})\> \omega_{ik}
\>\beta_i\>g(\theta_{ijk}), \end{equation}

\begin{equation} \label{eq:Tersoff2} g(\theta_{ijk}) = 1 + \frac{c_i^2}{d_i^2} -
\frac{c_i^2}{d_i^2 + \left(h_i - cos(\theta_{ijk})\right)^2} . \end{equation}

\noindent where $\theta_{ijk}$ is the angle between the atomic pairs $ij$ and
$ik$ bonds. Altogether, there are 13 various parameters in the T-P: $ A_{ij}$,
$B_{ij}$, $\lambda_{ij}$, $\mu_{ij}$, $R_{ij}$, $S_{ij}$, $\chi_{ij}$, $n_{i}$,
$\omega_{ij}$, $\beta_{i}$, $c_{i}$, $d_{i}$, and $h_{i}$. The parameters depend
on the chemical nature of the atoms $i$ and $j$, and their specific values for
certain elements have to be established through the suitably performed fitting
procedures.

In the present study, we use the following parameterisations of the T-Ps: the
Lindsay and Broido parameters for the C-C interactions (they were originally
developed for graphene and carbon nanotubes \cite{Lindsay}); the Albe and
M\"oller parameters for the B-B and N-N interactions (first introduced for
description of the hexagonal boron nitride based systems \cite{Albe}); and the
parametrization of Kinaci \textit{et. al} for the heteroatomic C-B, C-N, and B-N
bonds (developed earlier for h-BN/graphene interfaces in lateral graphene/h-BN
junctions \cite{Kinaci}).

\subsection{VFF-MC simulation scheme} 
\label{Methods-VFF-MC} 
In order to find equilibrium distribution of atoms over the lattice and atomic geometry of
mono-layered hexagonal CBN alloys, we employ our own simulation technique that
we refer to as VFF-MC scheme \cite{Michal,AJ}. We use Monte Carlo (MC) algorithm
in Metropolis NVT ensemble to search the configurational space, and the
Conjugent Gradient (CG) method to optimize geometry of the investigated
structures. Computational details look as follows. Assuming certain
concentration of carbon atoms, we distribute them randomly over the hexagonal
lattice points according to the three patterns corresponding to the various
growth condition: (i) over B sites, which corresponds to N-rich conditions, (ii)
over the N sublattice, which corresponds to N-poor conditions, and (ii) evenly
over N and B sublattices, which corresponds to the conditions inducing the
growth of stoichiometric C$_x$(BN)$_{1-x}$ alloys. Then we perform Monte Carlo
steps by swapping a randomly chosen pair of atoms and generating a new configuration. 
If the swapping procedure leads to the lower energy of the system the resulting configuration 
is unconditionally accepted, if not the new configuration is accepted with probability
 $p = \exp(\frac{ - \Delta E}{k_B T} )$, where $\Delta E$ is the
difference of the total energies of the system before and after the swap, $k_B$ is
the Boltzmann constant, and $T$ is a temperature like parameter that determines
only the swap probabilities in the Metropolis algorithm (T = 300K is used in simulations 
performed in these studies). The simulation
continues for few millions of MC steps (however, the convergence is typically
reached in $2 \cdot 10^6$ MC steps).

 The supercells used to study CBN alloys
contain 1800 atoms, and have rectangular shape with both zigzag and armchair
edges ('right/left' and 'top/bottom' edge in the figures presented later on,
respectively). This gives us possibility to investigate not only periodic 2D
systems (which is realized by assuming periodic boundary conditions in both x
and y directions), but also 1D nanoribbons (periodicity only in one direction),
and 0D flakes (no periodicity assumed), however, we focus on the 2D layered
systems in the present paper.

\subsection{Formation energy of defects and alloys}

Having determined the alloy's equilibrium distribution of atoms and geometry, we
can calculate the formation energy $E_{f}$ (also called enthalpy of formation)
of C-alloyed h-BN systems  following the well established formula
\cite{Zunger-dHf},

\begin{equation} \label{eq:Efq} E_{f} = E_{tot}{\rm [CBN]} - E_{tot}{\rm [h-BN]}+
n_{\rm B} \mu_{\rm B} + n_{\rm N} \mu_{\rm N} - n_{\rm C} \mu_{\rm C},
\end{equation}

\noindent where $E_{tot}$[CBN] and $E_{tot}$[h-BN] are the total energies of CBN
alloy and pristine h-BN, respectively, $n_{\rm B}$ and $n_{\rm N}$ indicate the
number of B and N atoms, respectively, which are substituted by $n_{\rm C}$
carbon atoms. For example, for isolated carbon impurity substituted in place of
B atom, $n_{\rm B}$ = $n_{\rm C}$ = 1, and $n_{\rm N}$ = 0. Further, $\mu_{\rm
B}$, $\mu_{\rm N}$, and $\mu_{\rm C}$ are the chemical potentials of B, N, and
C, respectively.

The chemical potentials represent the Gibbs free energy that isolated atoms
exchange with the heat reservoir. Those variables should represent experimental
conditions of alloy growth (growth conditions). The maximal possible values of
the chemical potentials $\mu_{i}^0$ (with i = B, N, C) correspond to the most
stable phases of the elements. Therefore, in this study, $\mu_{\rm B}^0$ is
referenced to as energy per atom in $\alpha$-rhombohedral phase of borophene,
$\mu_{\rm N}^0$ is energy per atom of nitrogen molecule, and $\mu_{\rm C}^0$ is
carbon energy in monolayer graphene, and all energies have been calculated
employing Tersoff's potential. In the real growth conditions, the chemical
potential $\mu_i$ of element $i$ can differ from its maximal value $\mu_i^0$,
and this difference is commonly indicated as $\Delta \mu_i$. In thermodynamic
equilibrium \cite{Zunger-dHf}, the differences in chemical potentials for two
elements forming a chemical compound must be equal to the compound's enthalpy of
formation $\Delta H_{f}$. For B and N pair, this rule reads $\Delta \mu_{\rm B}
+ \Delta \mu_{\rm N} = \Delta H_{f}({\rm BN})$, where $\Delta H_{f}({\rm BN})$ is
the enthalpy of formation of BN pair in h-BN, $\Delta H_{f}$(BN) $= E_{f}({\rm
BN}) - \mu_{\rm B}^0 - \mu_{\rm N}^0$, with $E_{f}({\rm BN})$ being the total
energy of BN pair in h-BN. The values of the characteristic energies obtained
with Tersoff's potential employed in our studies are: $E_{f}({\rm BN})$ = -15.01
eV, $\mu_{\rm B}^0$ = -6.84 eV, $\mu_{\rm N}^0$ = -4.96 eV, and $\mu_{\rm C}^0$
= -7.98 eV, leading to the h-BN formation enthalpy of $\Delta H_{f}({\rm BN})$ =
-3.27 eV. This value agrees very well with formation enthalpies calculated in
the framework of DFT for cubic and hexagonal bulk BN, and being equal
respectively to -3.0 eV and -3.3 eV \cite{cBN-hBN_H}. Computing formation
energies of alloys, we will consider the N-rich conditions (i.e., when the
system is in equilibrium with N$_2$ atmosphere), with $\mu_{\rm N} = \mu_{\rm
N}^0$, which gives $\Delta \mu_{\rm N}$ = 0 and $\Delta \mu_{\rm B} =
E_{f}({\rm BN})$, and also N-poor (B-rich) conditions, when $\Delta \mu_{\rm B}$
= 0 and  $\Delta \mu_{\rm N}$ = $E_{f}({\rm BN})$.
Before we turn to the discussion of predictions obtained within VFF-MC scheme
for monolayer CBN alloys, let us present some test calculations for homovalent
and heterovalent dimers involving B, C, and N, as well as for isolated carbon
defects C$_{\rm B}$ and C$_{\rm N}$, i.e., with carbon substituted in place of B
and N, respectively. This should help to assess the quality of the Tersoff's
potential.
\subsection{Validation of parameters set for 2D C-B-N systems}
\label{Methods-tests} 
\begin{figure}[htb] \includegraphics[width=0.45 \textwidth]{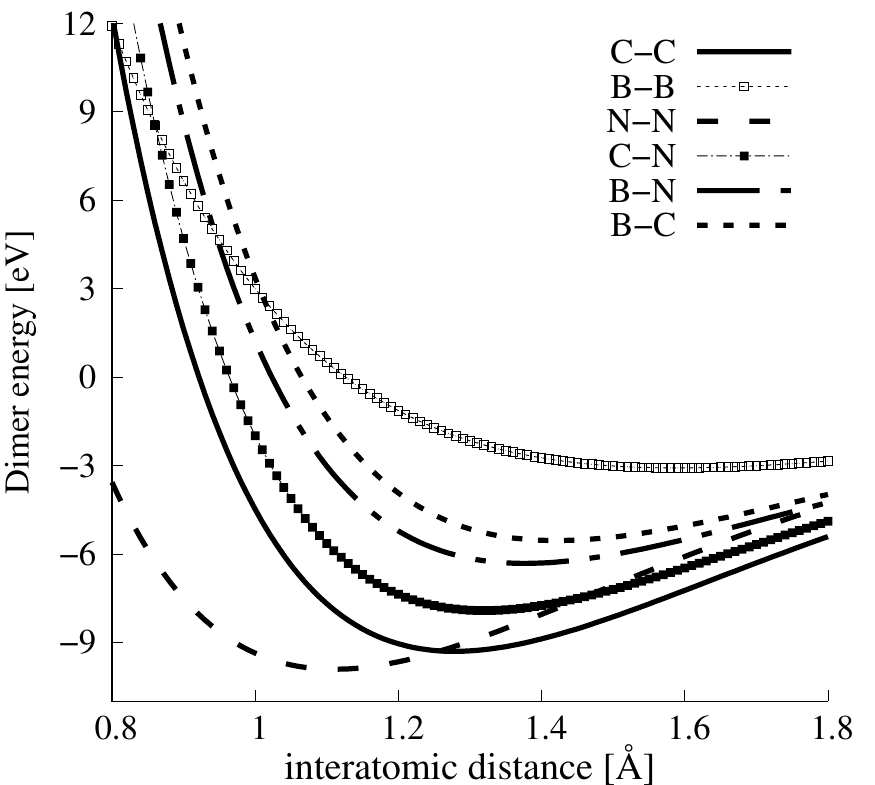}
\caption{Energy of dimer molecules as a function of interatomic distance.
	}
	\label{Fig2} \end{figure}

In the case of diatomic molecules, the Tersoff's potential
provides an analytic formula for equilibrium dimer bond length and dissociation
energy. Table~\ref{tab:Dimers} summarizes equilibrium bond lengths and energies
of all possible pairs of atoms forming dimer molecules out of atoms constituting
the CBN alloys (C-C, B-B, N-N, C-B, C-N, and B-N), whereas in Figure \ref{Fig2}, the
dimers' phase diagrams (the energy dependence on the distance between atoms) are
depicted. The comparison with available experimental data for dimers, 
corroborates that the Tersoff's potential provides overall reasonable description 
of the studied dimers.

\begin{threeparttable}[h]
	\label{tab:Dimers}
	\caption{Equilibrium bond lengths $d^{Ters.}_0$ and energies $E^{Ters.}$ of all dimers 
	corresponding to all possible bonds in CBN alloys calculated with Tersoff's potential,
	 compared to experimental values, $d^{Exp.}_0$ and $E^{Exp.}$, respectively.
	 }
	 
	\begin{tabular}[htbp]{@{}ccccc@{}}
		\toprule
		
		& $d^{Ters.}_0$ [\AA] & $E^{Ters.}$ [eV] & $d^{Exp.}_0$ [\AA] 
		& $E^{Exp.}$ [eV] \\
		\midrule
		
		\textbf{C-C}	& 1.28 	& -9.30	& 1.312 \tnote{(a)}	& -6.25 \tnote{(a)}	\\
						&		&		& 1.242 \tnote{(b)} &				   	\\
						&		&		& 1.240 \tnote{(c)} &				   	\\
						&		&		& 				 	& -6.244 \tnote{(d)} \\
		\hline	
		\textbf{B-B} 	& 1.59 	& -3.08 & 1.590 \tnote{(b), (c)}& 				\\
						&		&		& 	  				& -2.84 \tnote{(d)} \\	
		\hline
		\textbf{N-N}   	& 1.11  & -9.91 & 1.098 \tnote{(a)} & -9.763\tnote{(a)}	\\
						&		&		& 1.116 \tnote{(b)}	&	 \\						
						&		&		& 1.108 \tnote{(b)}	&	 \\						
		\hline	
		\textbf{C-B}   	& 1.42  & -5.54 & 					& -4.597 \tnote{(d)}\\
		\hline		
		\textbf{C-N}   	& 1.32  & -7.92 & 1.177 \tnote{(a)} & -7.721\tnote{(a)}	\\
						&		&		& 1.172 \tnote{(b)}	&				   	\\
		\hline		
		\textbf{B-N}   	& 1.37  & -6.32 & 1.281\tnote{(b)} 	& 					\\
		\bottomrule
		
	\end{tabular}
	\begin{tablenotes}\footnotesize
		\item (a) Ref. \cite{Di2}, (b) Ref. \cite{Di3}, (c) Ref. \cite{Di1}, 
		(d) Ref. \cite{Di5}. 	\end{tablenotes}
	
	\label{tab:Dimers}
\end{threeparttable}

As a next step of Tersoff's potential testing, we performed calculations for
C$_{\rm B}$ and C$_{\rm N}$ point defects. For these calculations, relatively
small supercells containing 200 atoms have been employed. The lattice relaxation
around impurity has been performed with fairly standard accuracy for forces,
which magnitudes in all directions have been required to be smaller than 0.01
eV/\AA. Our calculations for C$_{\rm B}$ impurity show that the
nearest-neighbors C-N bond length is equal to 1.43 \AA\ and compare very
favorably with {\it ab-initio} calculations in the DFT framework that predict
this distance to be equal to 1.409 \AA\ \cite{PRB86} for bulk hexagonal BN. For
C$_{\rm N}$ defect, we observe the distance between C-B atoms equal to 1.55
\AA\, in very good record with the DFT calculations that predicted this bond
length to be equal to 1.509 \AA\ \cite{PRB86}. In the case of N-rich conditions,
the formation enthalpies obtained on basis of Tersoff's potential are equal to
-3.19 eV and 6.80 eV, for C$_{\rm B}$ and C$_{\rm N}$ defects, respectively. In
qualitative agreement with {\it ab initio} calculations, our studies predict
that C$_{\rm B}$ defect has the lowest formation enthalpy in h-BN grown in
N-rich conditions. However, the obtained magnitudes of the formation enthalpies
for the C$_{\rm B}$ and C$_{\rm N}$ defects differ considerably from the DFT
based values reported in the literature. Namely, for monolayer h-BN grown in
N-rich conditions, the formation enthalpy of C$_{\rm B}$ was calculated to be
1.8 eV (with PBE exchange-correlation functional) and 1.95 eV (with HSE
exchange-correlation functional) \cite{PRL-h-BN}, whereas the formation enthalpy
for C$_{\rm N}$ defect was found to be 4.1 eV (PBE) or 4.3 eV (HSE)
\cite{PRL-h-BN}. Similarly, for hexagonal bulk BN, the formation enthalpies of
C$_{\rm B}$ and C$_{\rm N}$ impurities in N-rich growth conditions were found to
be equal to 1.7 eV and 4.3 eV \cite{PRB86}, respectively. Also the very new DFT
results for bulk hexagonal BN give the formation enthalpies of C$_{\rm B}$ and
C$_{\rm N}$ defects as 1.5 eV and 4.2 eV \cite{vdW-defects}, respectively. We
ascribe these discrepancies to the nature of Tersoff's potential, where charge
transfer effects are not taken into account, and, plausibly, they might be
relevant in the presence of heterovalent C-N and C-B chemical bonds.

To summarize, the Tersoff's potential satisfactorily describes geometry and
energetics of C, B, and N mutual dimers and carbon defects in h-BN,
corroborating that the VFF-MC scheme based on Tersoff's potential  may provide
valuable predictions for morphology, ordering, and energetic stability of CBN
alloys. However, before we turn to the discussion of this issue, we would like
to provide a short description of the tight-binding formalism for the low-energy
electronic structure of CBN alloys, which definitely increase our understanding
of these systems.

\subsection{Tight Binding approach for Density of States} 
\label{TB-theory} 
For
the description of the electronic structure of CBN alloys, we use simple
empirical tight-binding (T-B) theory considering only $\pi$-type orbitals. We consider
just first neighbor hopping characterized by single hopping parameter $t$.

\begin{equation} H_{TB} =\sum_i\epsilon_i|i><i|+t_0\sum_{<i,j>}|i><j| ,
\label{eq:HTB} \end{equation} \noindent where $|i>$ are $\pi$-states centered on
hexagonal lattice sites. We would like to stress that such simple T-B method
describes very well the low-energy spectrum around $K$-point in graphene and
h-BN \cite{dos1,dos2}. Numerous parameterizations of the T-B Hamiltonian in 
Eq.\ref{eq:HTB} are used in calculations of electronic structure
for CBN systems. In our calculations, we employ the 
parameters obtained through fitting the T-B electronic structure of h-BN  
to the predictions of the density functional theory \cite{TBPARAM}, and specifically 
$t_0$ = -2.16 eV, and the on-site energies of -2.55 eV, 0.0, and 2.46 eV, 
for nitrogen, carbon, and boron, respectively \cite{TBPARAM}. 
We have also used another set of T-B parameters \cite{TBparam2} to calculate 
the density of states for few test structures and observed that it leads only to 
insignificant changes.   

The T-B Hamiltonian of the size
$N_{lat} \times N_{lat}$, where $N_{lat}$ is the number of lattice points
considered within the VFF-MC formalism, can be easily diagonalized by any method 
for lattices considered in this paper. The projected density of states
(PDOS) on the state $|i>$ and density of states (DOS) are defined as follows

\begin{equation} \label{eq:pdos-dos}
D_i(\epsilon)=\sum_{\alpha}|<i|\alpha>|^2\delta(\epsilon-\epsilon_{\alpha}),
\end{equation} \begin{equation}
D(\epsilon)=\sum_{\alpha}\delta(\epsilon-\epsilon_{\alpha}). \end{equation}
\noindent For practical computations, we substitute delta $\delta$ function by
the Gaussian function $\tilde{\delta}$ of a given width $\sigma$
\begin{equation} \label{eq:delta}
\tilde{\delta}(\epsilon-\epsilon_\alpha)=\frac{1}{\sqrt2\pi\sigma}
\exp\big(-\frac{(\epsilon-\epsilon_{\alpha})^2}{2\sigma^2}\big). \end{equation}.

\section{Results} \label{RESULTS}

Now we turn to the discussion of the morphology, ordering, stability, and
electronic structure of monolayer honeycomb CBN alloys, which emerges from the
calculations within the developed VFF-MC + T-B scheme. In our calculation the
content of carbon atoms does not exceed 50\%. As it was described in the
introduction, the morphology of alloys depends very strongly on the growth
conditions. Here, we would like to present results of our calculations performed
in such a way that they mimic various growth conditions.

\subsection{CBN alloys in N-rich growth conditions} 
\label{RESULTS_C-hBN}

\begin{figure}[h!tb]%
\centering
\includegraphics[width=.48\linewidth]{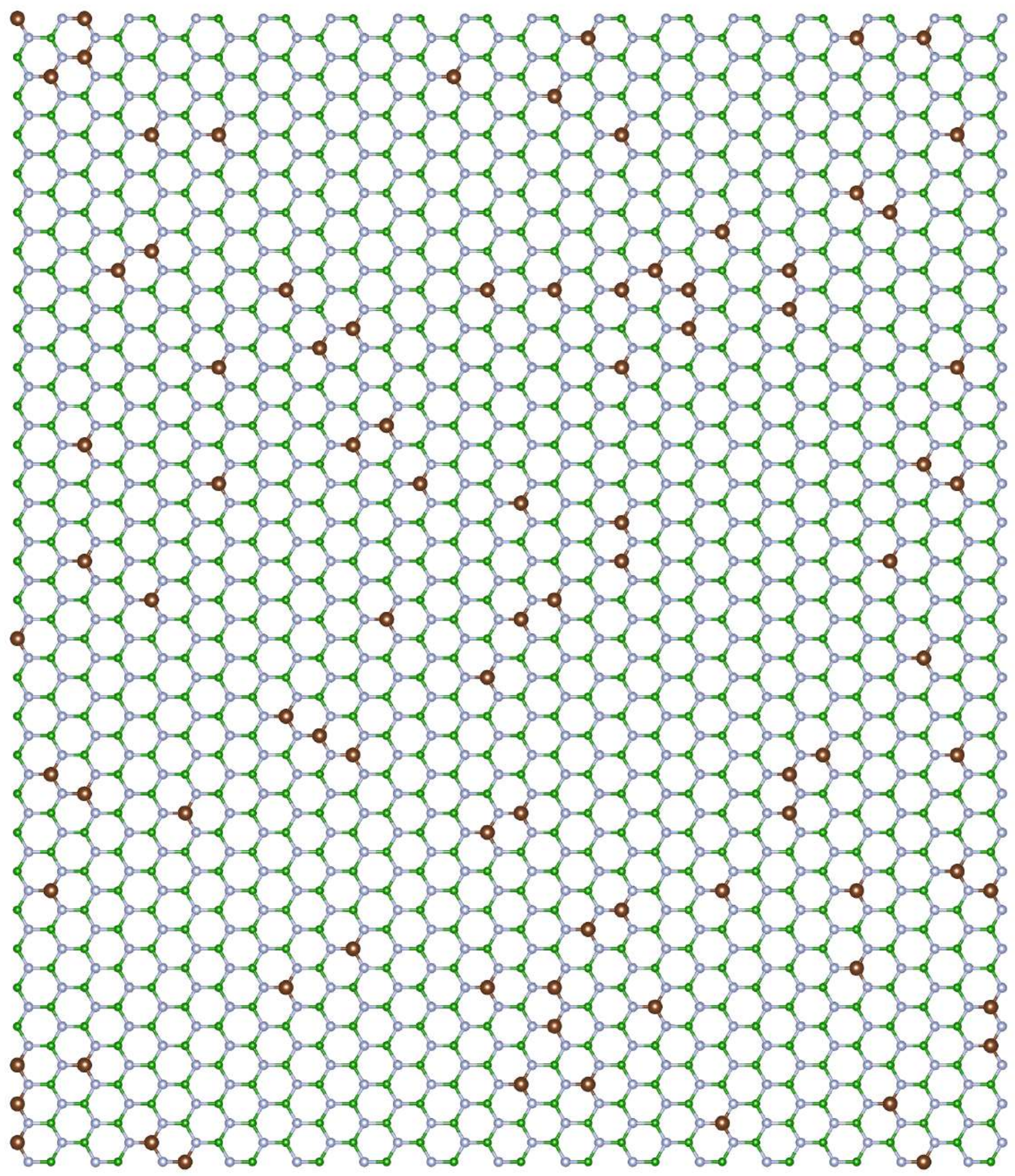}

	\caption{Distribution of C atoms within h-BN hexagonal lattice for the
	system with 10\% of C substituted initially randomly on the B sublattice only, 
	C$_{0.1}$B$_{0.4}$N$_{0.5}$ (N-rich). Green dots denote boron atoms, grey - nitrogen
	atoms and brown - carbon atoms (sizes of dots do not reflect atomic radii).
	}
	\label{POS-N-rich} \end{figure}

We have performed calculations for carbon concentrations ranging up to 30\%.
Figure~\ref{POS-N-rich} depicts the distribution of C atoms at exemplary
concentration of 10\% in CBN alloy after VFF-MC simulation.

\begin{figure}[h!] \includegraphics[width=0.4 \textwidth]{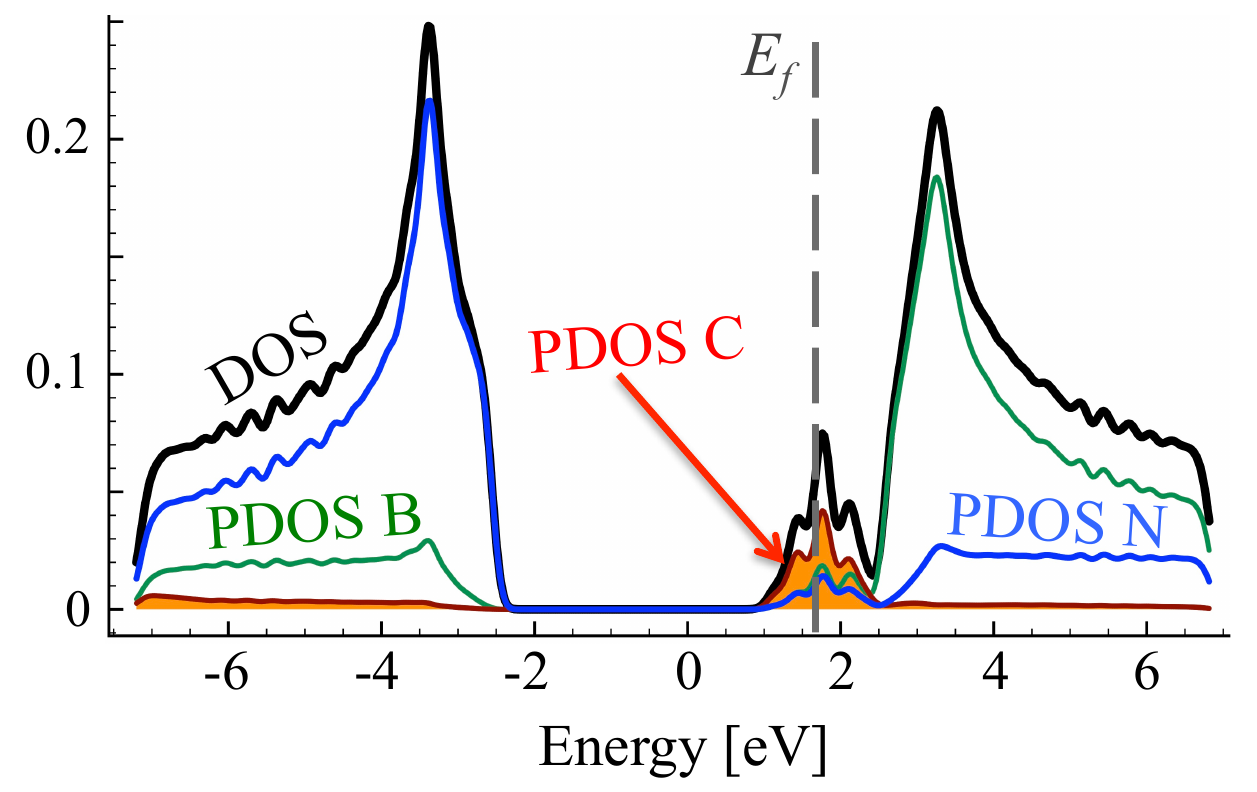}
\caption{TDOS and PDOS (projected on types of atoms) of ordered
C$_{0.1}$B$_{0.4}$N$_{0.5}$ system. Fermi level at $E_F$ = 1.78 eV is marked with
vertical dashed line.
	}
	\label{T-PDOS_10ConB} \end{figure}

As one can see, in the equilibrium structure C dopants are distributed
quasi-randomly within B sublattice, strongly indicating that C-N bonds are
rather energetically favorable, and no clustering happens in N-rich growth
mode. The formation energy of a CBN alloy, as calculated according to
Equation~\ref{eq:Efq}, is very close to sum of formation energies of isolated
C$_{\rm B}$ defects, i.e., $\Delta H^f_{N-rich} ( C
: h-BN ) /n_{C} = -3.19$ eV, where $n_{\rm C}$ is the number of C atoms
in the system, and exhibits extremely weak dependence on the concentration of C
atoms. Figure~\ref{T-PDOS_10ConB} depicts density of states for optimized
geometry of  C$_{0.1}$B$_{0.4}$N$_{0.5}$ alloy. One can observe impurity band
originating from carbon states around the Fermi energy, i.e., the typical donor
like band characteristic for carbon substituting boron. Note that in the case of
C atoms distributed over the B sublattice, the initial and optimized atomic
configurations exhibit very similar DOS.

\begin{figure}[h!] \includegraphics[width=0.4 \textwidth]{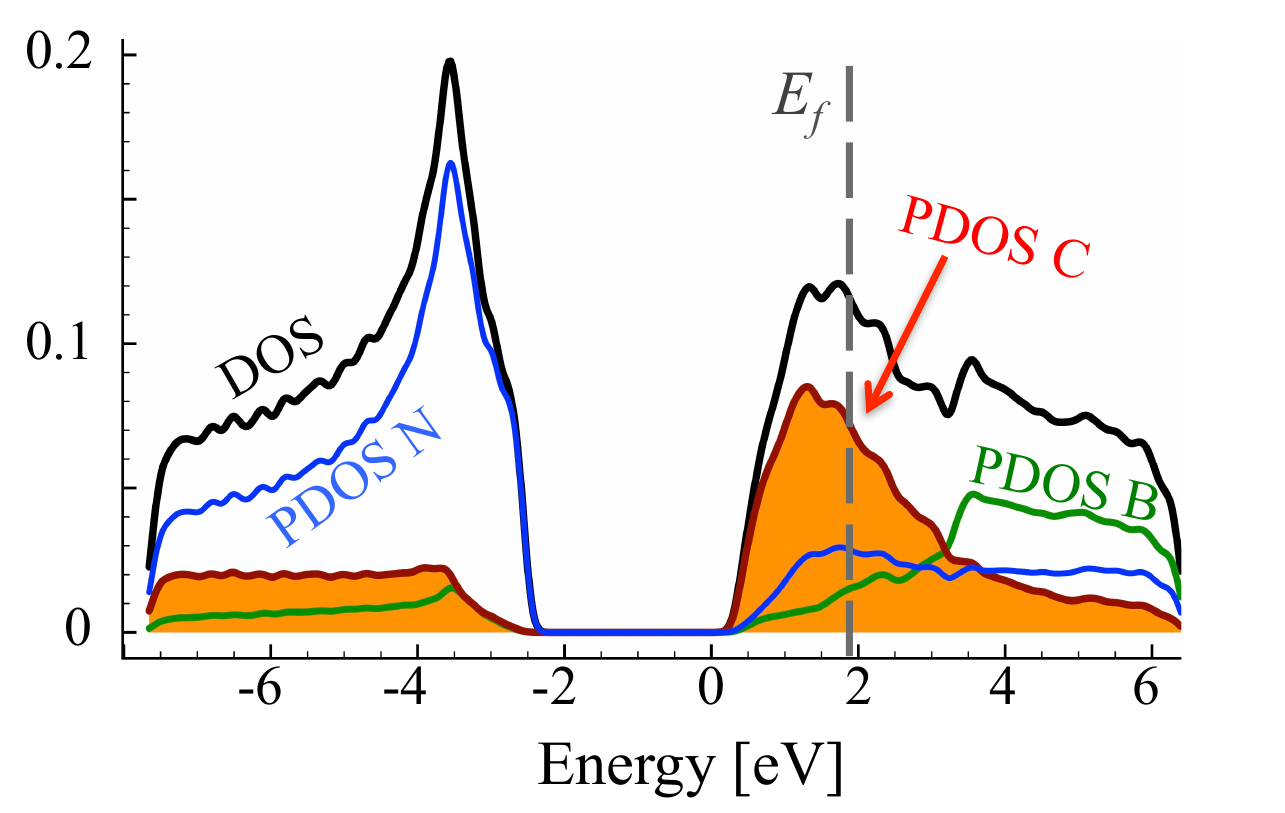}
\caption{TDOS and PDOS (projected on types of atoms) of ordered
C$_{0.3}$B$_{0.2}$N$_{0.5}$ system. Fermi energy at $E_F$ = 1.98 eV is marked with
vertical dashed line.
	}
	\label{T-PDOS_30ConB} 
\end{figure}

Even in the case of 30\% of B substituted by C atoms, there is no clustering, but
the density of states is significantly modified, especially in the region of
conduction band, as shown in~\ref{T-PDOS_30ConB}, and the Fermi level is shifted
towards higher energies.

\subsection{CBN alloys in N-poor growth conditions } \label{RESULTS_C-hBN}

\begin{figure}[h!tb]%
	 \centering

	\includegraphics*[width=.48\linewidth]{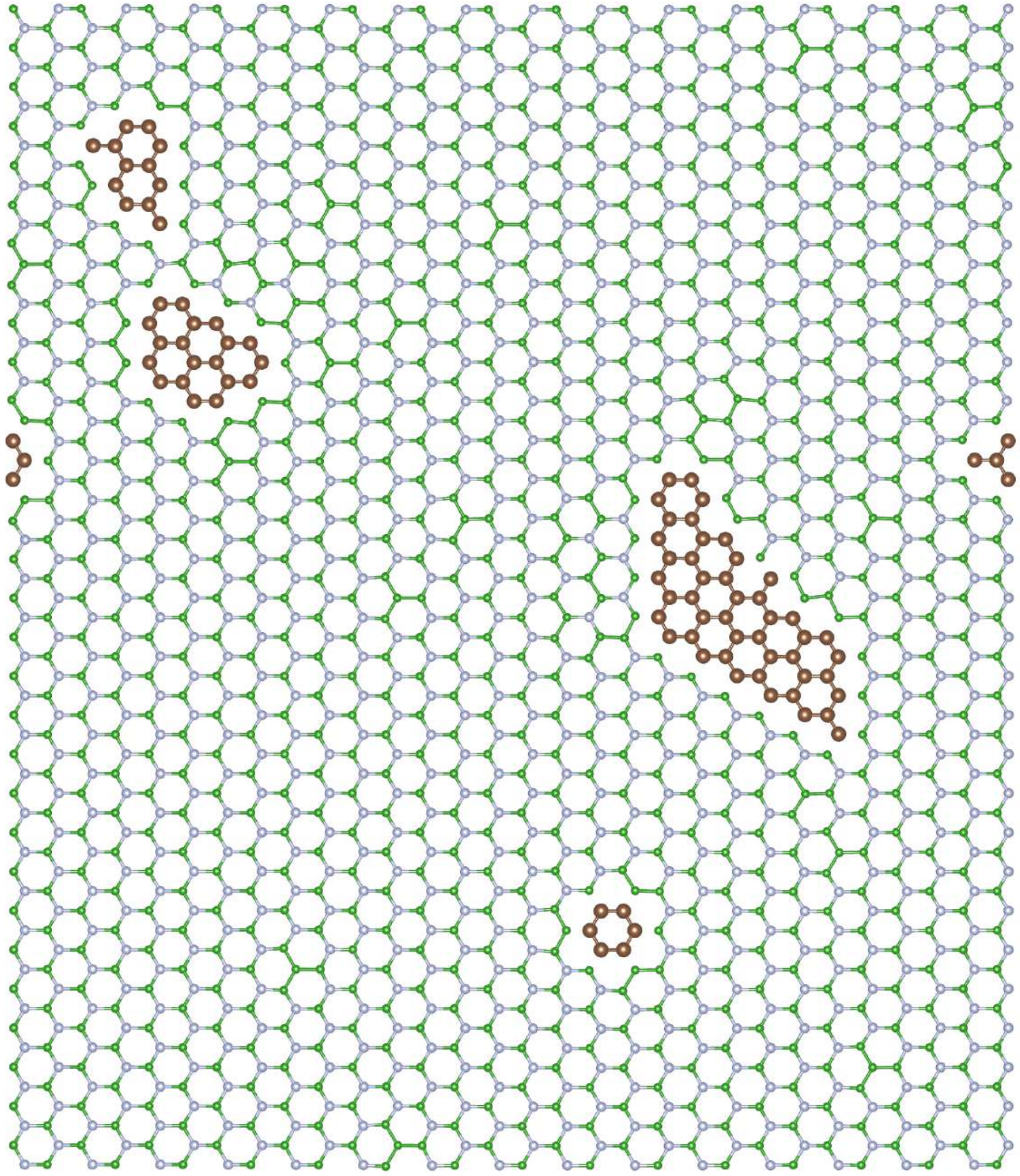} \caption{Distribution of C
	atoms within h-BN hexagonal lattice for the system with 10\% of C
	substituted initially only on N sublattice, i.e., C$_{0.1}$ B$_{0.5}$
	N$_{0.4}$ alloy. Green dots denote boron atoms, grey - nitrogen atoms and
	brown - carbon atoms (sizes of dots do not reflect atomic radii).
	}
	\label{POS-N-poor} \end{figure}

Now we turn to the case of CBN alloys grown in the N-poor conditions. It
corresponds to the situation when carbon atoms are substituted initially on
nitrogen sites (randomly), and the number of N atoms in the system is reduced. The
distribution of atoms in the energetically optimized configuration for the
carbon concentration of 10\% is depicted in figure \ref{POS-N-poor}.

\begin{figure}[h!] \includegraphics[width=0.4
\textwidth]{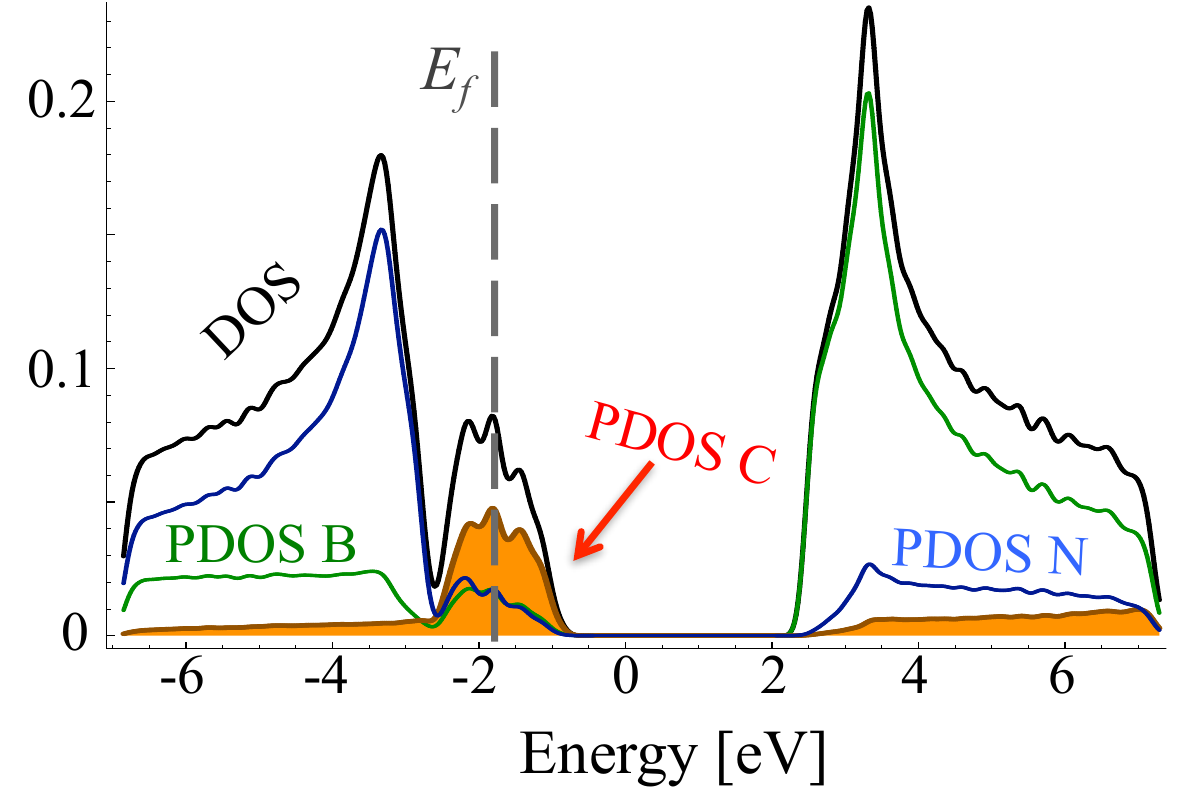} \caption{TDOS and PDOS (projected on types
of atoms) of C$_{0.1}$B$_{0.5}$N$_{0.4}$ system with randomly distributed C
atoms (within N sublattice). Fermi energy $E_F$ = -1.85 eV is marked with
vertical dashed line.
	}
	\label{T-PDOS_10ConN_rand} \end{figure}

\begin{figure}[h!] \includegraphics[width=0.4 \textwidth]{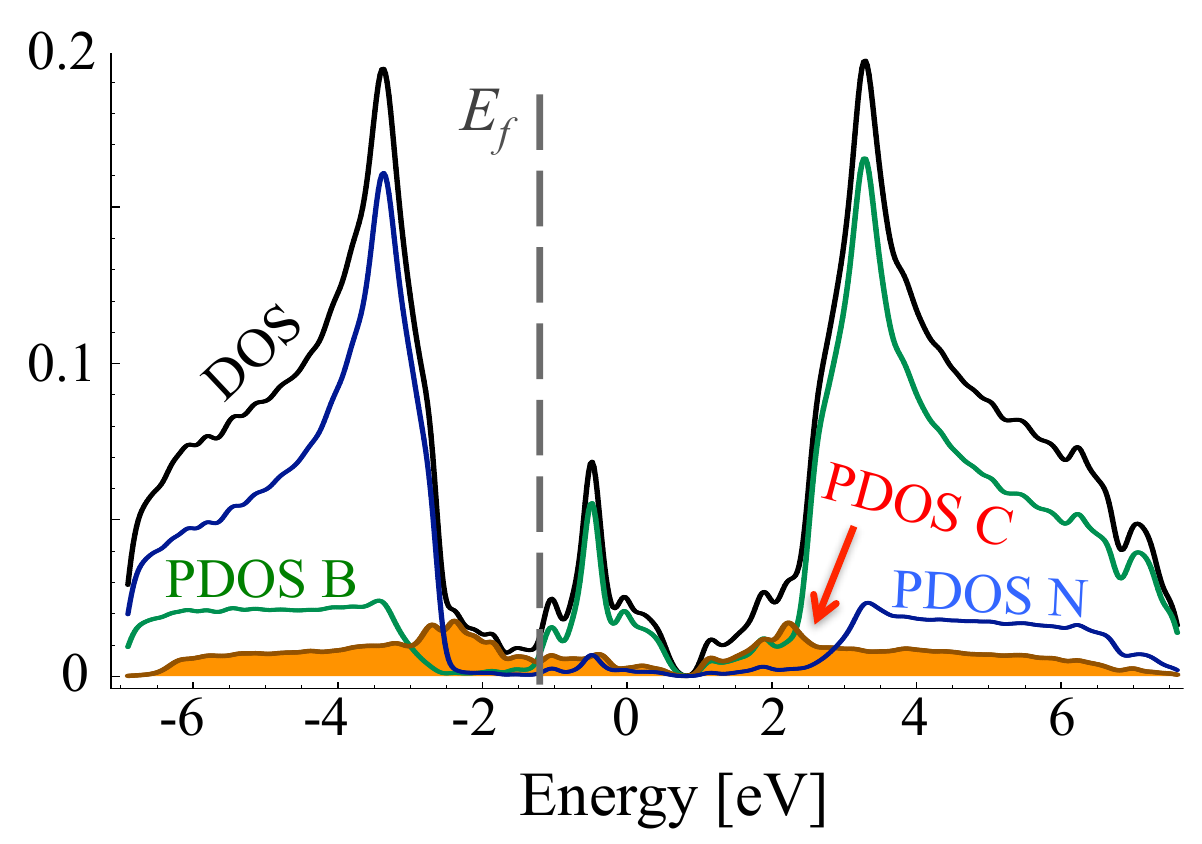}
\caption{TDOS and PDOS (projected on types of atoms) of ordered
C$_{0.1}$B$_{0.5}$N$_{0.4}$ system. Fermi energy $E_F$ = -1.23 eV is marked with
vertical dashed line.
	}
	\label{T-PDOS_10ConN_MC} \end{figure}

\begin{figure}[h!] \includegraphics[width=0.4 \textwidth]{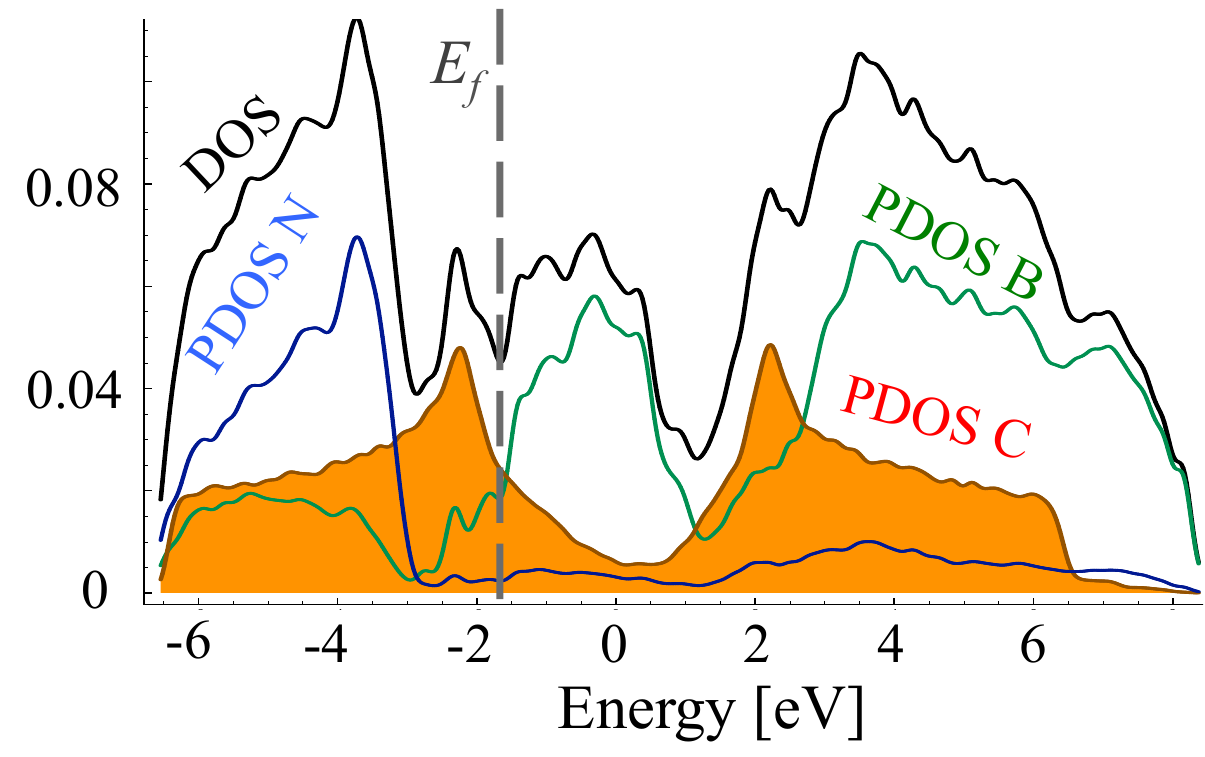}
\caption{TDOS and PDOS (projected on types of atoms) of ordered
C$_{0.3}$B$_{0.5}$N$_{0.2}$ system. Fermi energy $E_F$ = -1.64 eV is marked with
vertical dashed line.
	}
	\label{T-PDOS_30ConN} \end{figure}

It can be clearly seen that we have different behavior of the alloys in N-rich
(as described above) and N-poor conditions. In N-rich conditions, C dopants are
distributed quasi-randomly within B sublattice, while in N-poor case -
clustering of C atoms is observed for all investigated carbon concentrations.
The formation energy in systems with C substituted on N sublattice is very high
$\Delta H^f_{\rm N-poor} ({\rm C:h-BN})/n_{\rm C} > 2.5$ eV in all cases, with values
slightly higher for smaller concentrations of C. Thus we can conclude that in
N-rich growth conditions, there is possible to synthesize C-B-N alloy with
quasi-random distribution of C atoms, and such alloy should be more stable than
in the case of conditions with deficiency of N. It is clearly seen that after
the VFF-MC simulation carbon atoms form clusters of graphene 
in C$_{0.1}$B$_{0.5}$N$_{0.4}$ alloy, and also some B-B bonds appear. This causes
significant differences in DOS for structures: (i) with initial random
distribution of C over the N sublattice, and (ii) with final energetically
optimized morphology. This is depicted in Figures \ref{T-PDOS_10ConN_rand} and
\ref{T-PDOS_10ConN_MC}. With C atoms randomly distributed over N sublattice, one
observes states in DOS corresponding to individual C$_{\rm N}$ impurities,
whereas the energetically favorable structure exhibits following features in
its electronic structure, such as widen owing to the C-C interaction carbon
PDOS, and appearance of states of boron origin in the band gap of h-BN. This
effect is even more pronounced for C concentration of 30\%, as illustrated in
Figure \ref{T-PDOS_30ConN}, where C states seem to form graphene like DOS, and B
states largely modify area of h-BN gap.

The cases discussed so far reflect two extreme cases that can  emerge during
growth of C-doped h-BN systems, for example, within a CVD process. In the
following section, we present the result for intermediate case, i.e., when C
atoms are distributed equally between B and N sublattices. This corresponds to
the situation when we have proper concentrations of gases providing both N and B
to the system in equal amounts.

\subsection{Ordering and density of states in CBN stoichiometric alloys} \label{RESULTS-CBN}

\begin{figure}[htb]%
\begin{overpic}[width=0.19\textwidth]{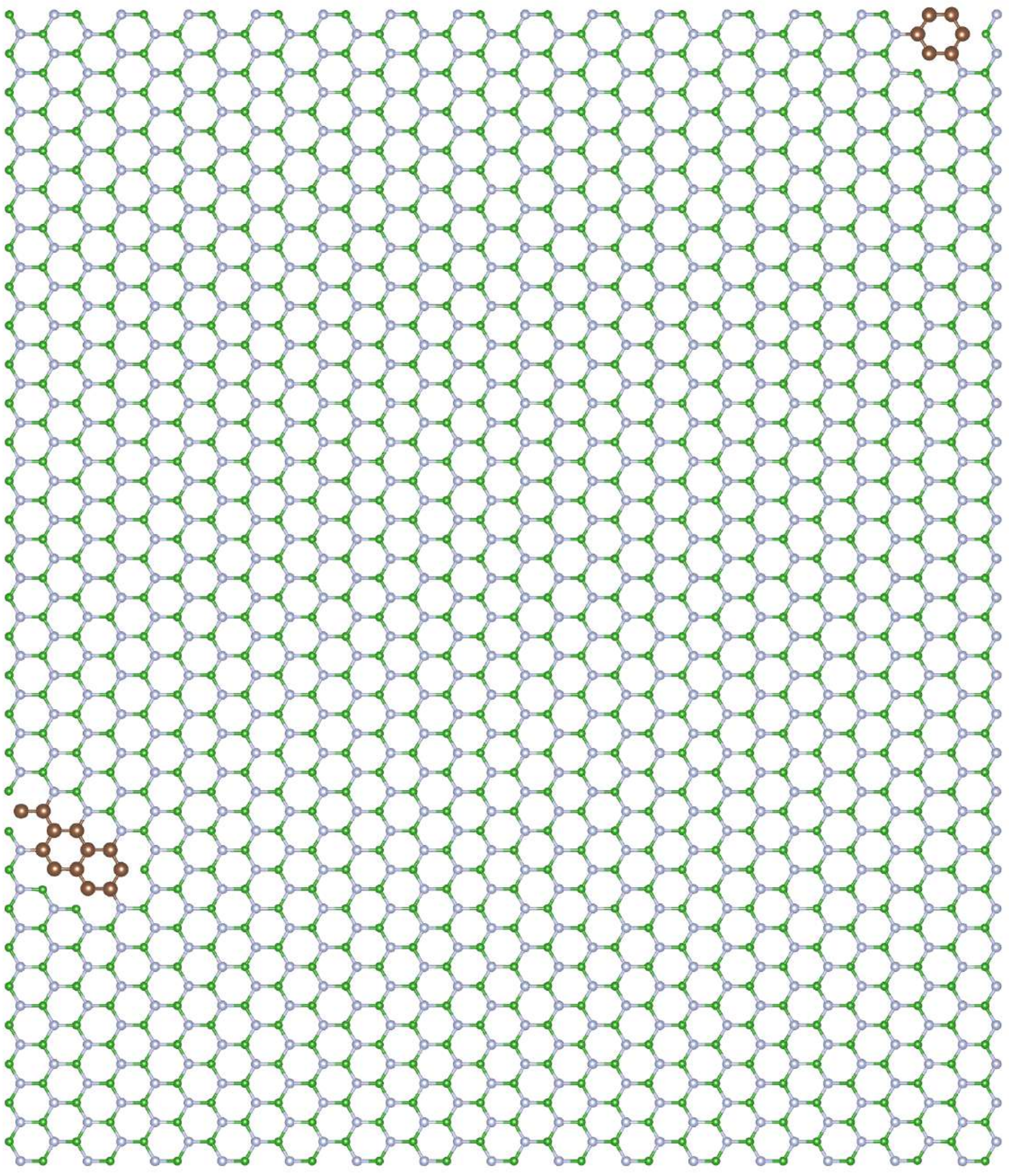}
 	\put (30,3) {\bf{ 1 \% C}}
\end{overpic}
\begin{overpic}[width=0.19\textwidth]{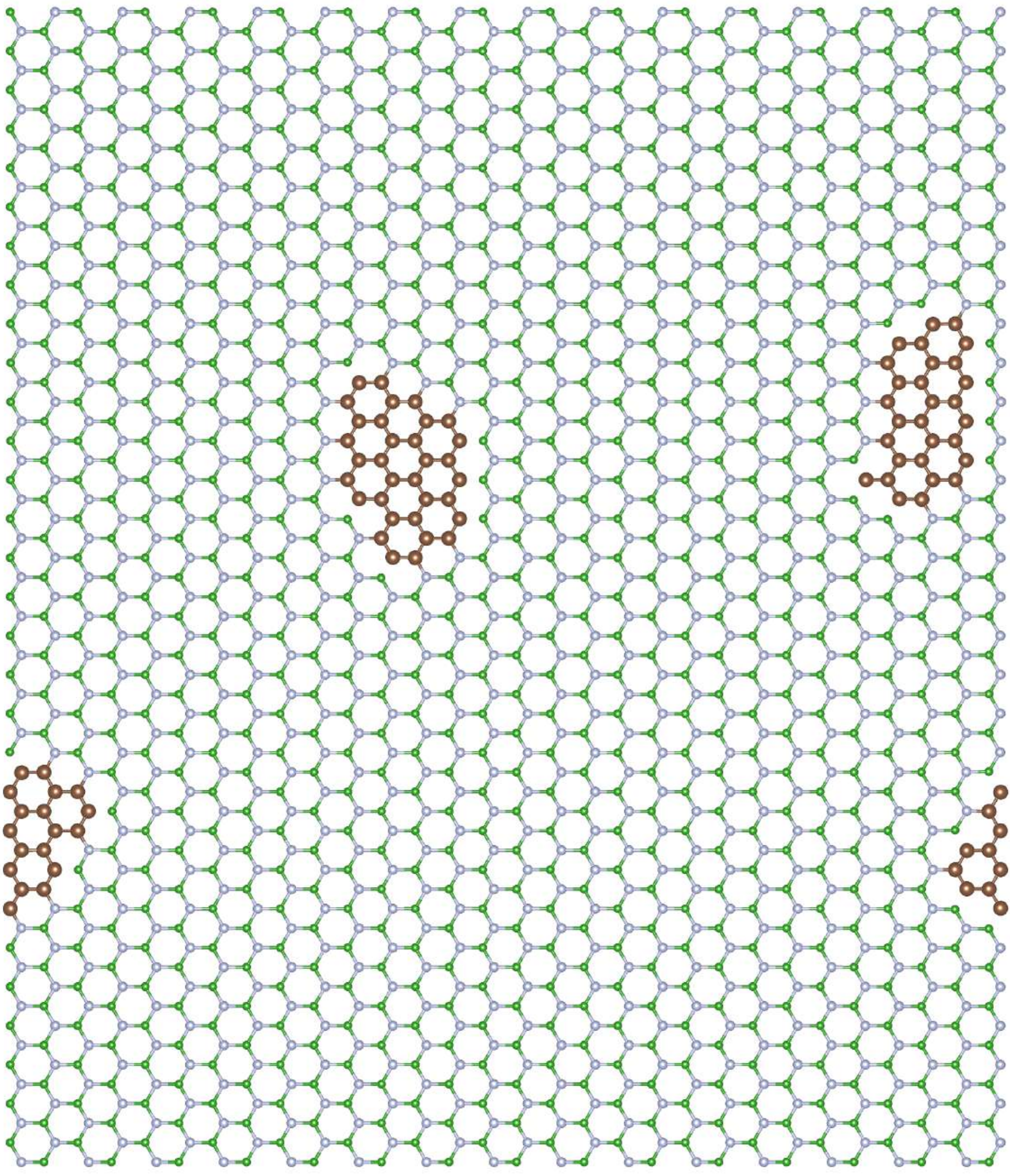} 
	\put (30,3){\bf{ 5\% C}}
\end{overpic}
\begin{overpic}[width=0.19\textwidth]{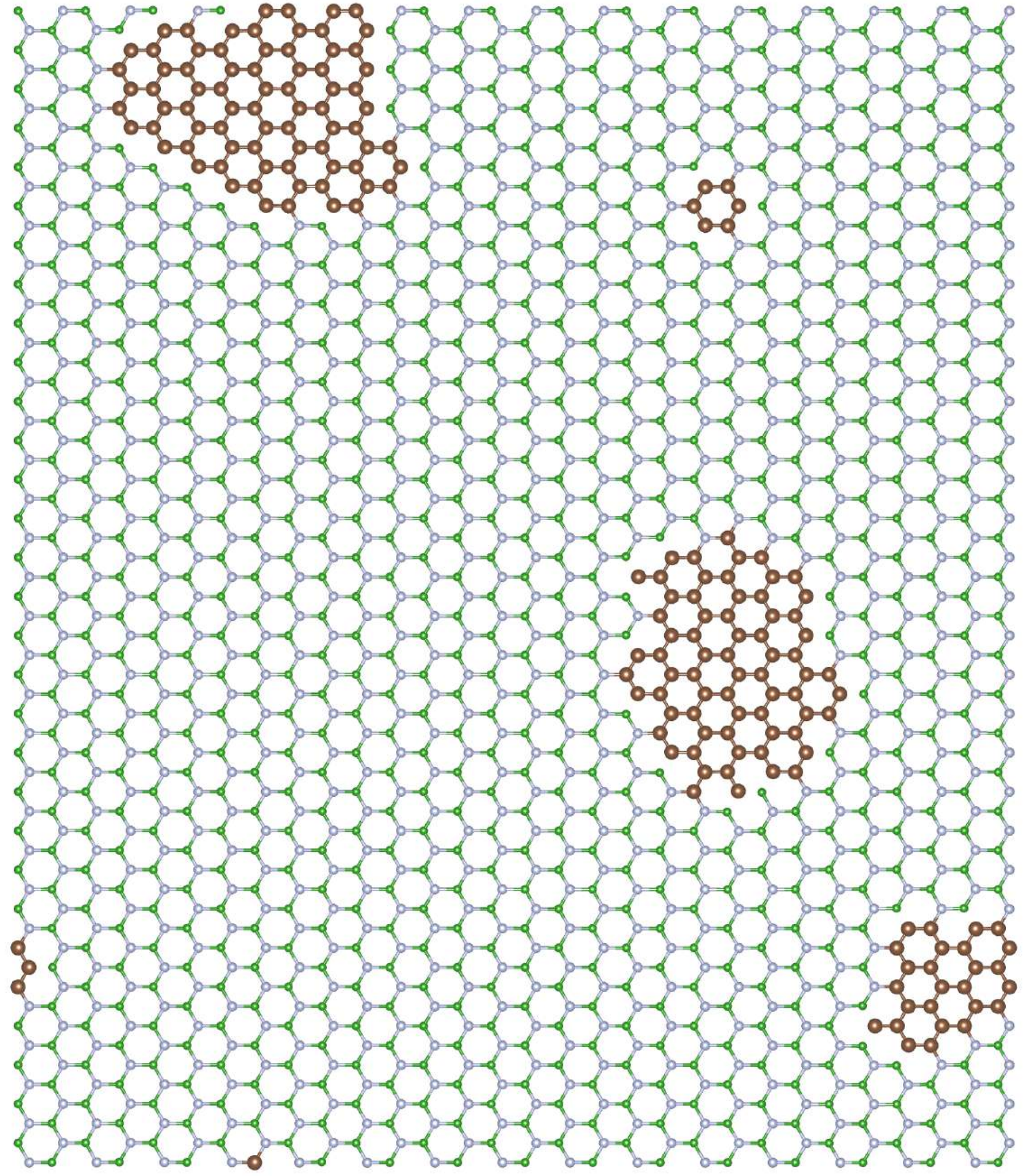}
	\put (30,3) {\bf{ 10\% C}} 
\end{overpic}
\begin{overpic}[width=0.19\textwidth]{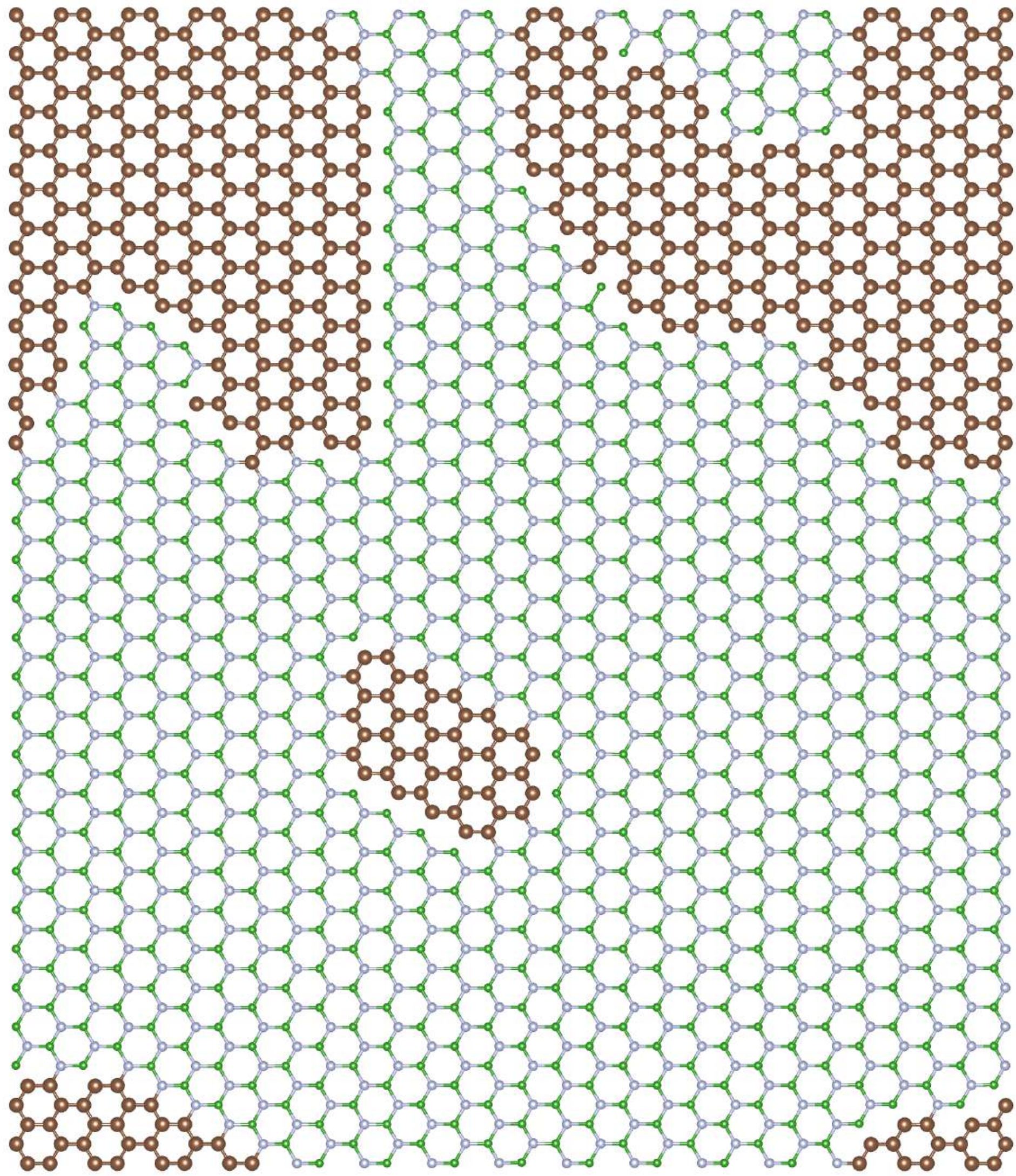} 
	\put (30,3) {\bf{ 30\% C}}
\end{overpic} 
\begin{overpic}[width=0.19\textwidth]{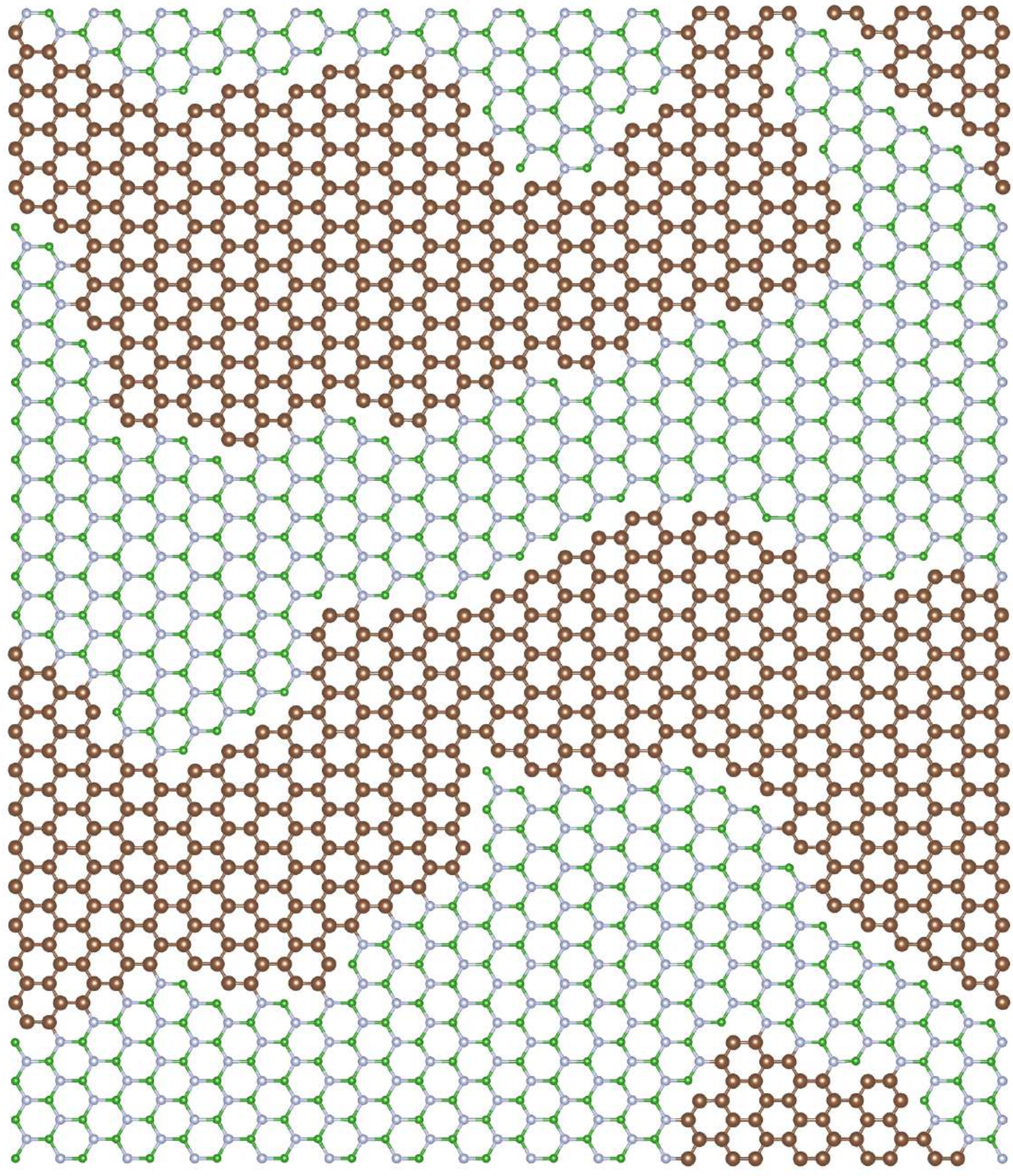}
	 \put (30,3){\bf{50\% C}}
\end{overpic} 
\caption{Distribution of C atoms within h-BN hexagonal lattice for various concentrations of carbon in the C$_x$(BN)$_{1-x}$ alloys. Green, grey, and brown dots indicate boron, nitrogen, and carbon atoms, respectively. The sizes of dots do not reflect atomic radii.} 
\label{POS-0-50-C} 
\end{figure}

\begin{figure}[htb]
\begin{overpic}[width=0.235\textwidth,height=3.6cm]{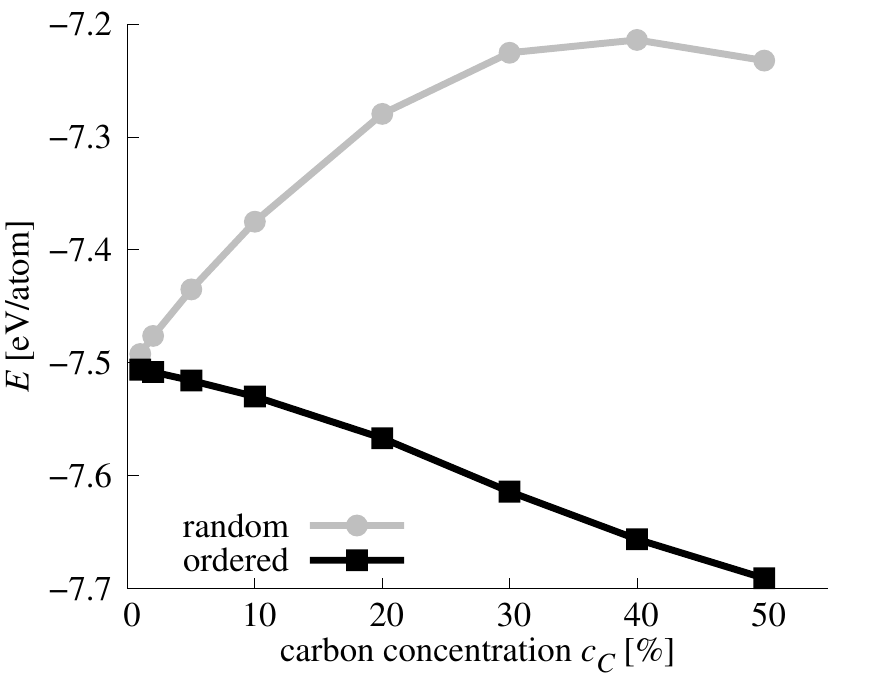}
	 \put(-3,83) {a)} 
\end{overpic} 
\begin{overpic}[width=0.235\textwidth, height=3.6cm]{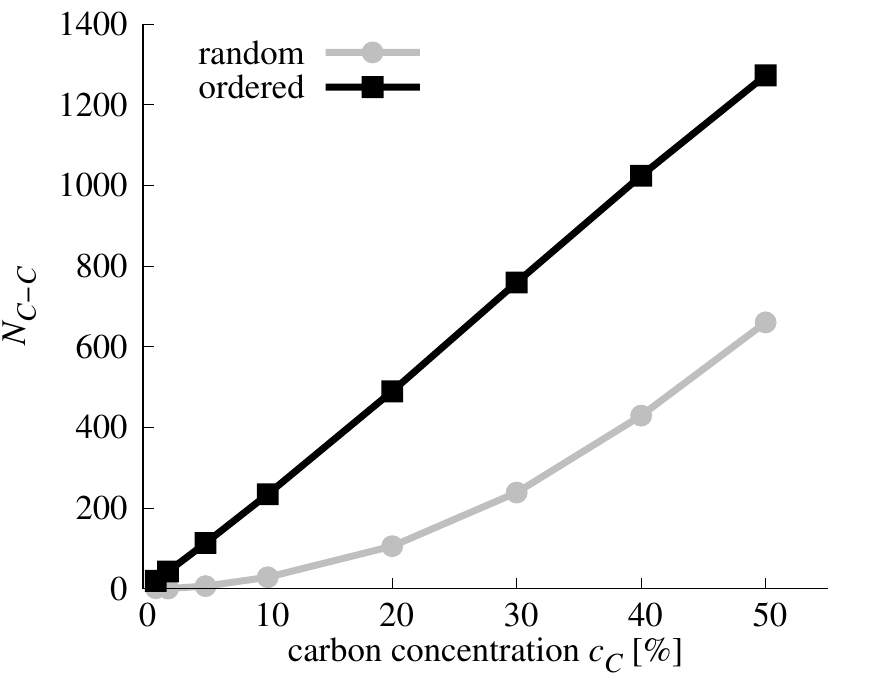} 
	\put (-3,83) {b)} 
\end{overpic}
\caption{Comparison of a) energy/atom and b) of number of C-C bonds for
C$_{x}$(BN)$_{1-x}$ alloys as function of concentration of C atoms, in random alloys and in systems with short range order (after MC-VFF simulation). } 
\label{ANALYS-prepostMC}
\end{figure}

We have performed simulations of stoichiometric C$_x$B$_y$N$_z$ alloys, i.e.,
when $y=z=0.5-0.5x$. Henceforward, we will refer to them as C$_x$(BN)$_{1-x}$ 
systems. We have performed VFF-MC simulation to obtain equilibrium distribution
of atoms for x in the range of 1-50\%. The resulting distributions of carbon atoms
over the lattice  are presented in Figure~\ref{POS-0-50-C} for the exemplary concentrations 
cases of x = 1, 5, 10, 30, 50 \%. It is clearly seen that, for all
concentrations of carbon, separation of graphene (GR) and h-BN phases appears.
This result is in accordance with experimental results, that frequently report
clustering of GR and h-BN in ternary alloys \cite{Natcom,Uddin,ChemPLett,NatureMat}. 
On the basis of our theoretical investigation,
this result can be explained by the fact that C-C and B-N bonds are much more
energetically favorable in comparison to C-N and particularly to C-B bonds,
thus during MC simulation system tends to minimize number of the latter. We have
compared energy per atom and number of C-C bonds in the system at
non-equilibrium (i.e., random in our case) and in the system corresponding to the thermodynamic 
equilibrium (i.e., after MC procedure) for all investigated concentrations of C. They are
shown in Figure~\ref{ANALYS-prepostMC}. 

\begin{figure}[h!]
\includegraphics[width=0.45  \textwidth]{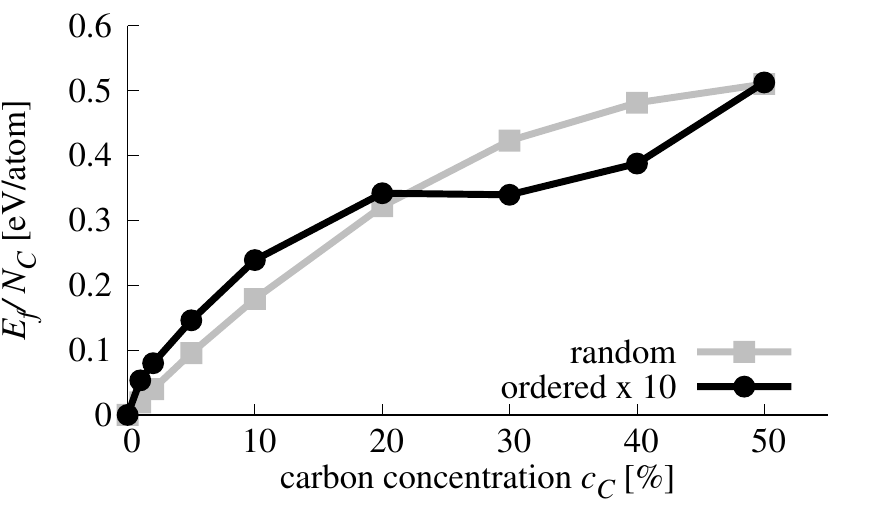}
\caption{Formation energy for  C$_{x}$(BN)$_{1-x}$ alloys as a function of
concentration of C atoms, in random systems and systems with the short range order (i.e., after Monte Carlo configuration optimization). Note that the values for alloy with short range order are multiplied by factor 10.}
\label{FormationE_stochio} 
\end{figure}

As can be seen in Figure~\ref{ANALYS-prepostMC} b), the number of C-C bonds in
equilibrium structure deviates strongly from number of these bonds in the case
of randomly distributed C atoms. This happens in the whole range of
concentrations. The formation energies of the stoichiometric C$_x$(BN)$_{1-x}$ alloys, 
both with random and optimized distribution of atoms   
are depicted in Figure~\ref{FormationE_stochio}. It is worth mentioning that in the case 
of these alloys $n_{\rm B}$ = $n_{\rm N}$ = $n_{\rm C}/2$, and the alloy formation energy 
in Eq.(\ref{eq:Efq}) may be expressed in terms of the carbon concentration 
$x$ = $n_C/N_{at}$ (with $N_{at}$ being the total number of atoms in the system) as follows
 
\begin{equation}  \label{eq:DEb} \Delta E_{b}(x) = E_{\rm CBN}(x) 
-(1 - x)E_{\rm h-BN} - xE_{\rm GR},
\end{equation}

\noindent where the $\Delta E_{b}(x)$ is the so-called \textit{mixing energy} 
commonly used in the literature as the indication of the alloy formation 
energy, and $E_{\rm CBN}(x)$, $E_{\rm h-BN}$, and $E_{\rm GR}$ are 
the energies per atom  for C$_x$(BN)$_{1-x}$ alloy, h-BN, and graphene, respectively. 
As one can see, the alloy formation energies for the  C$_x$(BN)$_{1-x}$ alloys with 
energetically optimized distribution of C atoms (i.e., exhibiting short range order) 
are generally an order of magnitude lower (maximal energy is of order 0.04 eV) 
than the alloy formation energies 
for random alloys (maximal energy is of order 0.4 eV). 
However, they are both positive, which indicates the tendency 
for the phase separation between h-BN and graphene.  
In the case of alloy with short range order, one observes also a minimum 
in $\Delta E_{b}(x)$ for $x$ around 30\%, whereas for the random alloys, $\Delta E_{b}(x)$ 
exhibits monotonic (nearly parabolic) dependence on the carbon concentration $x$.                                   

Qualitatively, the similar behavior of alloy formation energies for 
C$_x$(BN)$_{1-x}$ systems was observed in Refs.\cite{rev2_2,JAComp708}, 
where stability of nanoribbons with zig-zag and armchair interfaces 
between h-BN and graphene was investigated. The maxima of alloy formation energies 
lie there by 0.15 eV and 0.25 eV for structures with zig-zag and armchair interfaces, 
respectively. Also $\Delta E_{b}(x)$ reaches there some plateau for carbon concentration 
above 30\%, however, there is so pronounced minimum of $\Delta E_{b}(x)$ as observed 
in the present study for C$_x$(BN)$_{1-x}$ alloys 
with short range order see Figure \ref{FormationE_stochio}. 

Having calculated the mixing energy $\Delta E_{b}(x)$, it is very easy to calculate 
free energy defined as 
\begin{equation}
\label{FreeEn}
F(T,x) = \Delta E_{b}(x) - T S(x), 
\end{equation} 
\begin{equation}
\label{sx}
S(x) = -k_B [xlnx + (1-x)ln(1-x)], 
\end{equation}
\noindent where $k_B$ is the Boltzmann's constant, and $S(x)$ entropy of mixing. 
Such procedure was performed in the Refs. \cite{PRB79,PRB95} and the critical 
temperatures for the miscibility of carbon in h-BN were determined. Our estimations lead 
to the conclusion that the full miscibility could be achieved only above the melting 
temperature of h-BN. Taking into account that we (in analogy to Refs.\cite{PRB79,PRB95}) 
do not consider vibrational entropy contribution to $F(T,x)$ and also configurational 
part of the entropy should be modified in the case of the short range order present 
in the alloy, we will perform the discussion of the phase diagrams for C$_x$B$_y$N$_z$ 
alloys elsewhere.  

\begin{figure}[h!]
\includegraphics[width=0.45
\textwidth]{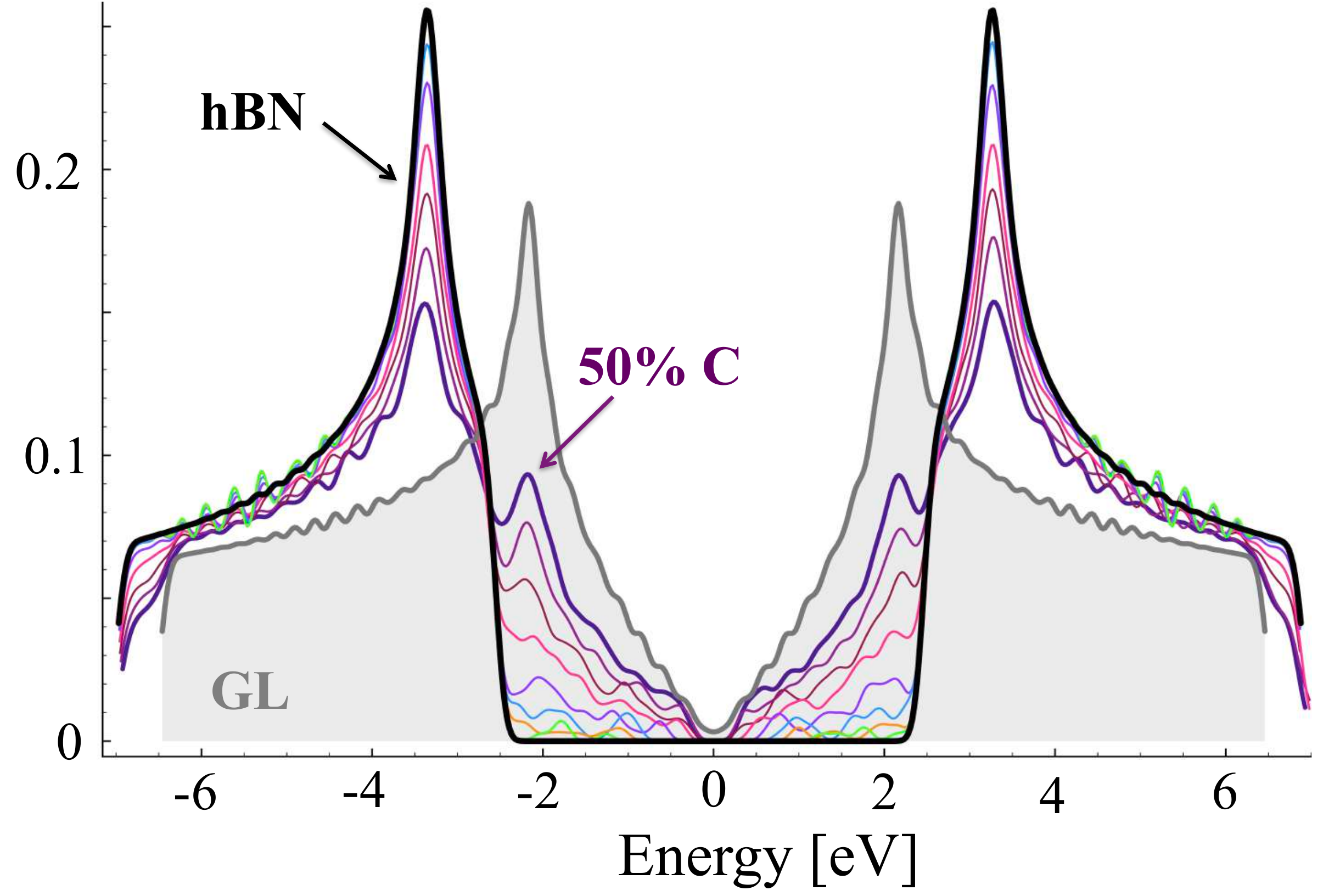} 
\caption{Total density of states for C$_{x}$(BN)$_{1-x}$ for several C concentrations up to 50\%}
\label{TDOS_C-hBN} 
\end{figure}

\begin{figure}[h!]
\includegraphics[width=0.45 \textwidth]{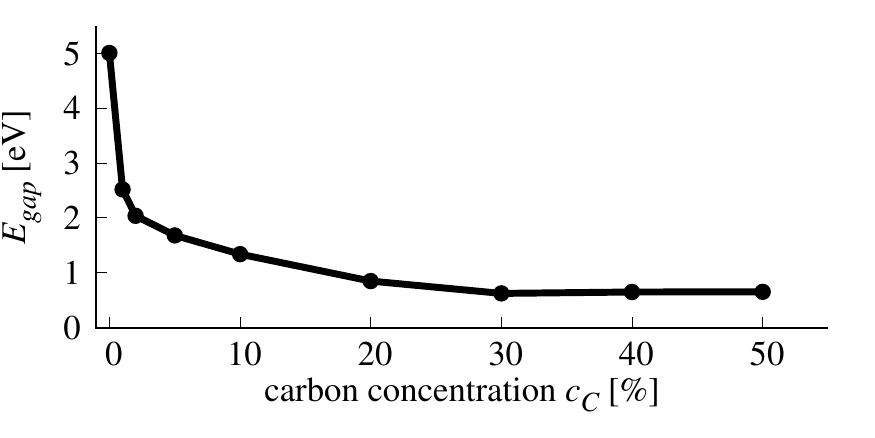}
\caption{Band-gap $E_{gap}$ of C$_{x}$(BN)$_{1-x}$ alloys as a function of
carbon concentration $c_C$ up to 50\% .}
\label{Egap_concC} 
\end{figure}

The results for density of states for stoichiometric compounds are presented in Figure~\ref{TDOS_C-hBN}, whereas, the dependence of band gap on C concentration is shown in Figure~\ref{Egap_concC}. The band gap decreases very fast with increasing concentration of C in the system, and is below 1 eV for
concentrations higher than 20\%, just exhibiting drop by 4 eV. This would
suggest possible way for tuning the band gap in CBN systems. Similar non-linear absolute
decrease of the band gap was previously observed in several experimental 
\cite{rev2_3,Uddin} and theoretical studies 
\cite{rev2_2,rev2_3,JAComp708,C2016}, however, the band gap bowing observed 
in the bulk h-BN \cite{Uddin} is much smaller than the value reported in this paper. In other studies, the decrease of energy gap in C$_x$(BN)$_{1-x}$ alloy with C concentration is very abrupt in full analogy to our findings.  

\begin{figure}[h!]
	\centering
	\includegraphics[width=0.4\textwidth]{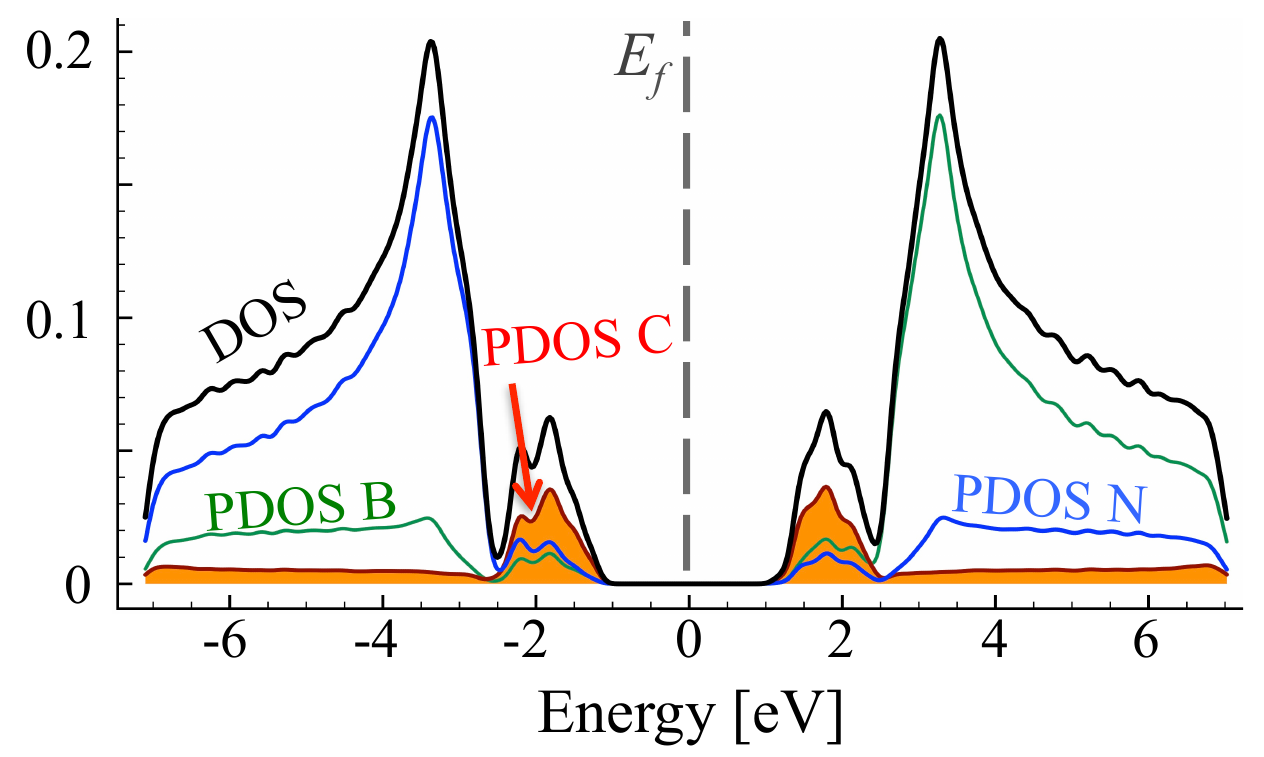}
	\caption{Total and projected DOS for C$_{0.1}$(BN)$_{0.9}$ alloy with non-equilibrium	random distribution of C atoms.}
	\label{T-PDOS_10C_rand} 
\end{figure}

\begin{figure}[h!]
	\centering
	\includegraphics[width=0.4\textwidth]{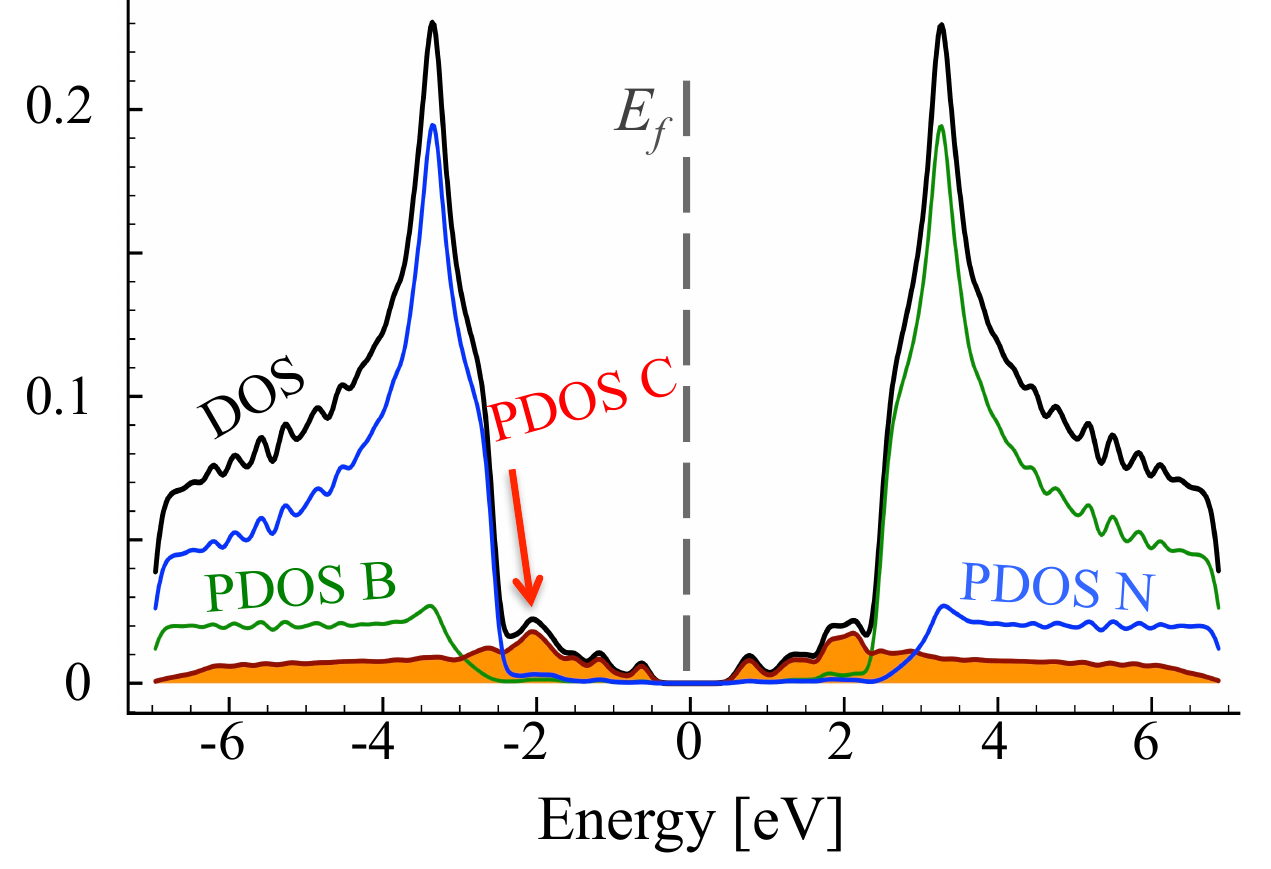} 
	\caption{Total and projected DOS for C$_{0.1}$(BN)$_{0.9}$ alloy with equilibrium morphology exhibiting segregation of graphene (GR) and h-BN domains.}
	\label{T-PDOS_10C_MC} 
\end{figure}

Figures~\ref{T-PDOS_10C_rand} and~\ref{T-PDOS_10C_MC} show TDOS and PDOS for
exemplary case of C$_{0.1}$(BN)$_{0.9}$ before and after MC simulation, i.e.,
for non-equilibrium random and equilibrium distribution of C atoms exhibiting short range 
order, respectively. One can notice significant difference of Projected Density of
States connected with carbon atoms. For random distribution of C, one can notice
significant contributions of carbon states in the area of h-BN band gap. They
are connected to not clustered C$_{\rm B}$ and C$_{\rm N}$ impurities in h-BN
lattice which act like donors and acceptors, respectively. The structure is
different in the system with short range order. Carbon states are more uniformly 
distributed, and the band-gap is much smaller due to the presence of graphene 
areas in the system.

\section{Summary} \label{Summary}

In the summary, we have performed calculations of morphology, ordering,
stability, and electronic structure of h-BN rich monolayer CBN alloys with
carbon concentrations below 50\%, paying particular attention to accounting for
the various growth conditions in experiments. In agreement with
some experimental and theoretical studies, we find that in the most cases h-BN
and graphene segregate building domains or precipitates. However, we predict
that the CBN alloys with the homogeneously distributed carbon atoms could only
form during growth process with the N-rich conditions. Such alloys have metallic
character with Fermi level lying in the carbon induced impurity band. In the
C$_x$(BN)$_{1-x}$ stoichiometric alloys, the band gap exhibits very fast non-linear 
decrease with growing carbon concentration.

\begin{acknowledgement} This work was supported by the National Science Centre
in Poland (NCN) through the grants PRELUDIUM (No. UMO-2017/25/N/ST3/00660) 
and OPUS-12 (UMO 2016/23/B/ST3/03567).
\end{acknowledgement}

\newpage

\listoffigures

\section*{Graphical Table of Contents\\} GTOC image: \begin{figure}[h]%
\includegraphics[width=4cm,height=4cm]{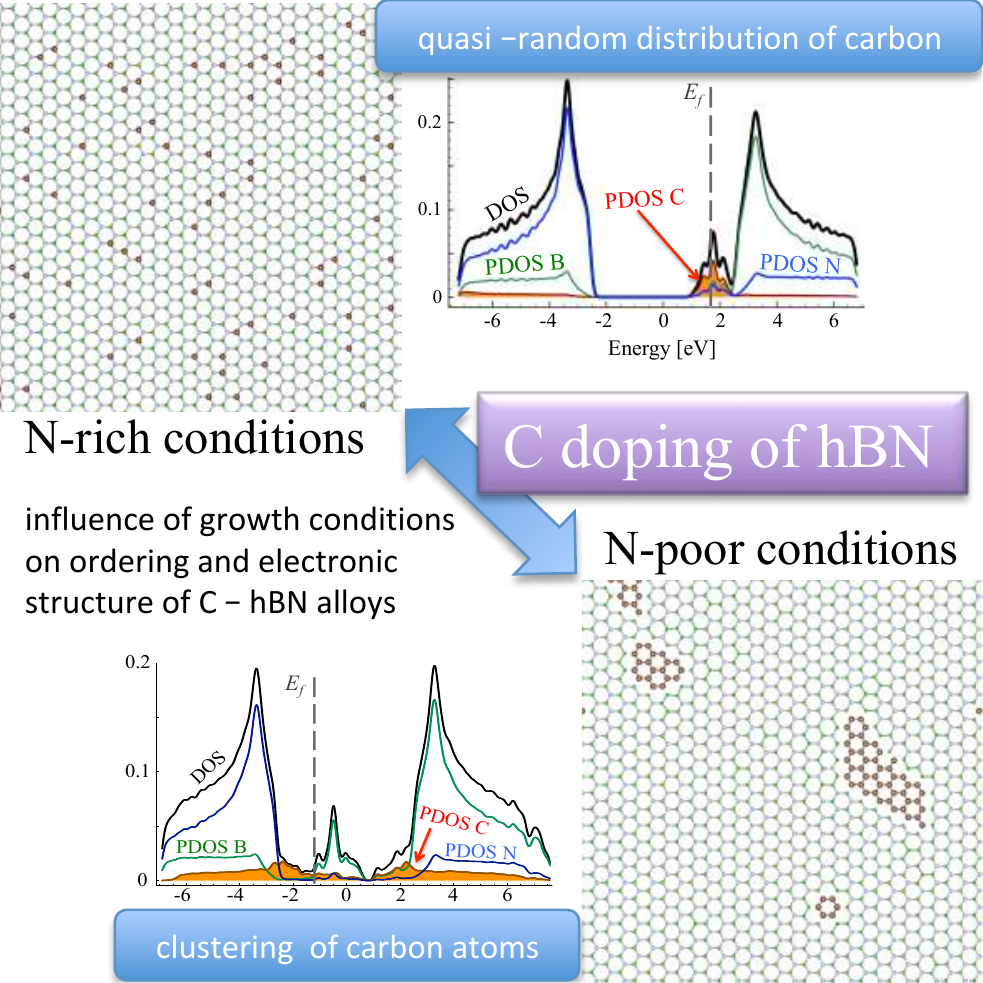} \caption*{ Your article will
be published with a Graphical Abstract in the table of contents. Please send a
suggestion for an image (preferably full colour, size 4 cm x 4 cm). It may be
specifically designed for the purpose, but should not show too many details or
consist of several parts. Enclose a short descriptive and popular text on the
general aim and value of your paper which may serve as an `appetizer' for the
readers (40--70 words, not a figure caption, not the abstract text).}
\label{GTOC} \end{figure}

\end{document}